\documentclass[12pt, preprint]{aastex63}
\usepackage{amsmath}
\usepackage{amssymb}
\usepackage{amsfonts}
\usepackage{graphicx}
\usepackage{color}
\usepackage{xcolor}
\usepackage{xfrac}
\usepackage[caption=false]{subfig}

\newcommand{\be}{\begin{equation}}
\newcommand{\ee}{\end{Fuation}}
\newcommand{\bea}{\begin{eqnarray}}
\newcommand{\eea}{\end{eqnarray}}

\begin{document}

\title{Probing Axial Symmetry Breaking in the Galaxy with {\it Gaia} Data Release 2}

\author[0000-0002-9785-914X]{Austin Hinkel}
\affiliation{Department of Physics and Astronomy, 
University of Kentucky, Lexington, KY 40506-0055}

\author[0000-0002-6166-5546]{Susan Gardner}
\affiliation{Department of Physics and Astronomy, 
University of Kentucky, Lexington, KY 40506-0055}

\author[0000-0002-9541-2678]{Brian Yanny}
\affiliation{Fermi National Accelerator Laboratory, Batavia, IL 60510}

\begin{abstract}

We study a set of solar neighborhood ($d < 3$ kpc) stars from Gaia Data Release 2 to determine azimuthal star count differences, i.e., left and right of the line from the Galactic center through the sun -- and compare these differences north and south. In this companion paper to \citet{GHY20}, we delineate our procedures to remove false asymmetries from sampling effects, incompleteness, and/or interloper populations, as this is crucial to tests of axisymmetry. Particularly, we have taken care to make appropriate selections of magnitude, color, in-plane Galactocentric radius and Galactic $|b|$ and $|z|$. We find that requiring parallax determinations of high precision induces sampling biases, so that we eschew such requirements and exclude, e.g., regions around the lines of sight to the Magellanic clouds, along with their mirror-image lines of sight, to ensure well-matched data sets. After making conservative cuts, we demonstrate the existence of azimuthal asymmetries, and find differences in those, north and south. These asymmetries give key insights into the nature and origins of the perturbations on Galactic matter, allowing us to assess the relative influence of the Magellanic Clouds (LMC \& SMC), the Galactic bar, and other masses on the Galactic mass distribution, as described in \citet{GHY20}. The asymmetry's radial dependence reveals variations that we attribute to the Galactic bar, and it changes sign at a radius of $(0.95 \pm 0.03)R_0$, with $R_0$ the Sun--Galactic-Center (GC) distance, to give us the first direct assessment of the outer Lindblad resonant radius. 

\end{abstract}


\section{Introduction}

Models of the Galaxy are often 
motivated by the assumption that it is isolated and thus has certain integrals of motion.
It is, moreover,
commonly regarded as a superposition of its disk, halo,  bulge, and bar components \citep{robin2003synthetic, robin2012stellar}. 
Each component $i$ can be modelled by a distribution
function $f_i$ \citep{binney2008GD, binney2012actions, bovy2013direct, piffl2014constraining, piffl2015bringing}, which, in steady-state, is characterized by its integrals of motion, as predicated by Jeans' theorem \citep{jeans1915}. 
It is useful to model $f_i(\mathbf{J})$, where $\mathbf{J}$ has 
as components the action integrals $J_r$, $J_\phi$, and $J_z$, in
radial, azimuthal, and vertical coordinates 
$r$, $\phi$, 
and $z$ 
with respect to the plane of the Galactic disk. Notably 
$J_\phi$ (or ``$L_z$") is the angular momentum 
about the symmetry axis of an axisymmetric disk. 
Each $f_i(\mathbf{J})$ is a supposed invariant under
the slow evolution of the Galaxy. 
Although the Galaxy has features that are notably axially
asymmetric, namely, the spiral arms and Galactic bar, it is
nevertheless the case that $f({\mathbf{J}})$ modelling \citep{binney2014galactic} gives
a very good description of the velocity distributions
observed by the RAVE survey. 
With the advent of {\it Gaia} DR2 data \citep{prusti2016gaia,lindegren2018gaia} it has become
possible to determine the structure of the dark halo 
from the stars alone \citep{binney2015distribution}, in that the
circular speed curve with the in-plane Galactic 
radius $R$ from the reconstructed Galactic potential 
is compatible with the circular speed from Cepheids
\citep{mroz2019rotation, binney2019modelling}. 

An integral of motion used to model the Galactic 
distribution function should also correspond to a continuous
symmetry, as the converse
of Noether’s Theorem \citep{noether1918} 
holds if the integral of motion is non-zero~\citep{olver2012applications}. 
Thus, in regions nominally described by just a disk and spheroidal halo, we expect the angular momentum about the $z$-axis, $L_{z}$, to be an integral of motion and thus axial symmetry should be manifest.  However, we know structures such as 
the Galactic bar and spiral arms do not exhibit axial symmetry, and there are also satellite galaxies which may influence the Milky Way.  It is thus interesting to assess the degree to which the galaxy is truly axisymmetric away from these known asymmetric sources, 
as doing so can provide insights into the validity of the above assumptions and help to 
determine which perturbers may be most relevant.  

\begin{figure}[ht!]
\begin{center}
\includegraphics[scale=0.5]{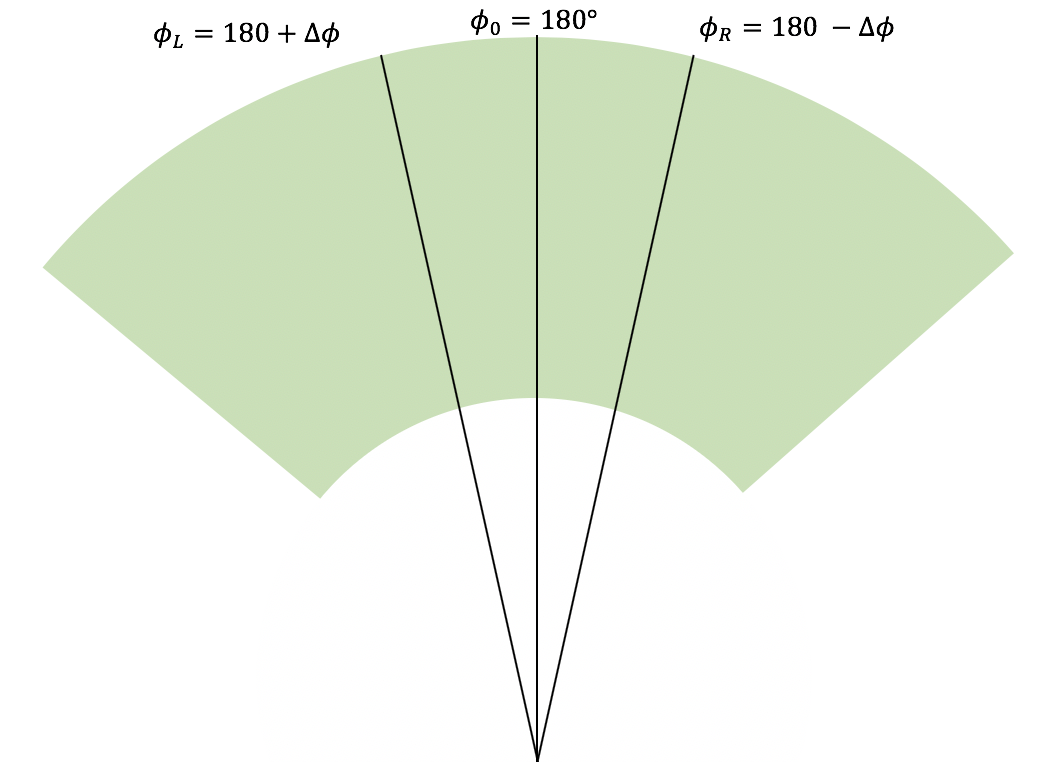}
\caption{A schematic depiction (i.e., not to scale) of the method of computing an axial asymmetry. 
To construct ${\cal A} (\phi)$, one counts the number of stars, $n_{\rm R}$, in a range $\delta \phi$ about an 
angle $\phi_{\rm R}$ to the right of the $\phi = 180^{\circ}$ line, that is, in a bin centered on $\phi_{\rm R}$, 
and then counts the number of stars, $n_{\rm L}$, in a 
bin of width $\delta \phi$ centered on the angle $\phi_{\rm L}$ 
to the left of the $\phi = 180^{\circ}$ line, chosen such that
$\phi_{\rm L} = 180^{\circ} + \Delta \phi$, where 
$\phi_{\rm R} = 180^{\circ} - \Delta \phi$ 
and $\Delta \phi > 0$. 
The asymmetry can be computed from 
$n_{\rm L,R}$ via Eq.~\ref{AsymmetryParam}.  Note we choose 
uniform bins in $\delta \phi$ across the region of interest,
so that we can assess $A(\phi)$ with $\phi$. 
}
\label{fig:schematic}
\end{center}
\end{figure}

In practice, to test axial symmetry, one simply needs to count the stars in a given range of galactocentric longitude, $\phi$, and then compare the counts of that bin with the bin corresponding to a galactocentric longitude range that has been reflected across the $\phi = 180^{\circ}$ line.  With the star counts of the left ($n_{L}$, $\phi > 180^{\circ}$) and right ($n_{R}$, $\phi < 180^{\circ}$) bins, one is able to compute the asymmetry, as defined in \citet{GHY20}:
\begin{equation}
    {\cal A} (\phi) = \frac{n_{L}(\phi) - n_{R} (\phi)}{n_{L}(\phi) + n_{R}(\phi)} ,
    \label{AsymmetryParam}
\end{equation}
where the functions $n_{L,R} (\phi)$ contain integrals over
the other coordinates as appropriate. 
Computing axial asymmetries for the Galaxy would, in theory, be possible with respect to any Galactocentric longitude, but due to observational constraints driven by the heliocentric nature of our $|b| > 30^{\circ}$ latitude cuts and the limited reach of the {\it Gaia} telescope, we use the $\phi = 180^{\circ}$ line as our baseline and limit our probe to galactocentric longitudes within $6^\circ$ of this baseline and within galactocentric in-plane distances of 
$7 < R < 9$ kpc, 
where the Sun is assumed to be at $R=R_0 = 8$ kpc, 
though we note the recent result of $8.178 \pm 0.026 $ kpc from 
\citet{abuter2019geometric}, 
and vertical distances of no more than 3 kpc above or below the plane.  A schematic depiction of the (in-plane projected) geometry involved in this process can be found in Fig.~\ref{fig:schematic}. If tests of axial symmetry
about other ``fold'' lines were ever to become practicable, our ability
to probe dynamical effects, such as from 
the presence of the Galactic bar, would
be greatly heightened. 

To our knowledge, a test of axial asymmetry in stars out of the 
 Galactic mid-plane region is first employed in \citet{GHY20}.  
 In this companion paper
 we carefully develop the data selection procedures that 
 make the studies of \citet{GHY20} possible. 
 There we conclude that significant asymmetries in the
 left versus right star counts combined with North and South sub-counts 
 clearly imply the influence of massive objects 
 which break the axial symmetry of 
 the Milky Way's stellar and overall matter distributions. 
 Comparing 
 the relative torques due to well-known ``perturber'' 
 candidates, we found
 the influence of the LMC to be dominant. However, due to uncertainties 
 in the mass distributions, including those 
 from unseen dark matter, we cannot 
 rule out that 
 some fraction of the mismatch in counts left and right and 
 north and south could be due to other structures, such as the 
 Sagittarius (Sgr) dwarf 
 tidal stream and the Galactic bar.  \citet{GHY20}
 was further able to 
 distinguish between oblate and prolate mass 
 distributions by looking at 
 not only ``global'' several degree scale (in $\phi$)  axial asymmetries, 
 but also how these asymmetries differed north and south of the 
 mid-plane. In \citet{GHY20} we considered the analysis volume 
 we develop in this paper, namely, the selection  $7 
 < R < 9$ kpc, $0.2 < |z| < 3$ kpc, $|b| > 30^\circ$ as a whole, along with brightness and color selections
 and particular sight-line exclusions. 
 Here, 
 we sub-divide the data further in an attempt to find out if the 
 LMC is responsible for the entirety of the asymmetries, and we 
 discover
 important asymmetric effects from the Galactic bar. In
 what follows we carefully 
 delineate our data selection and analysis procedures
  before turning to a discussion of the axial asymmetries, in 
  aggregate and in various subspaces of our analysis data set.

\section{Data Selection and Completeness Studies} \label{sec:dataselect}

The {\it Gaia} documentation and supporting publications \citep{luri2018gaia, bailer2015estimating} suggest using a Bayesian approach to estimate distances, particularly if the relative errors in parallax are large. 
We do not use such an approach here, as we do not need to know a precise distance to a star, just the correct azimuthal bin into which it falls. 
Thus we estimate the distance to each object as the inverse of its 
parallax, $\varpi$, throughout. 
Additionally, we do not see any evidence that the measured 
parallaxes are systematically correlated with a star's location
left or right of 
the $\phi = 180^{\circ}$ line. 

As noted by \citet{luri2018gaia}, using small relative parallax error quality cuts introduces artifical asymmetries, or biases, into 
a data set, because the parallax error is a function of position on the sky due to the {\it Gaia} scanning law's non-uniform 
sampling of different regions of the sky.  This would irreparably bias the study of symmetries over an appreciable extent of the galaxy, as shown in Fig.~\ref{fig:lbcuts} a).  Instead, we outline here a method for selecting a set of stars which appears to be free of any obvious biases (Fig.~\ref{fig:lbcuts}b) and is smoothly distributed in multiple parameter spaces, indicating that no chunks of data are missing or significantly afflicted by observational biases.  To wit, we use the photometric and astrometric data to effect cuts which preserve a smooth distribution of stars in G-band magnitude, $G_{\rm BP} - G_{\rm RP}$ color, in-plane radius from the 
GC, and height above the plane over a range of Galactocentric longitudes centered on the anti-center line, $\phi = 180^{\circ}$. The logic of the alternative method we propose is as follows: if there is an artificial 
bias in the {\it Gaia} data set, it would manifest itself as some aberration from the expected smoothly varying distributions in at least one parameter space.  In the subsections that follow, we outline each parameter space checked and use the plots generated in this study to motivate a maximal angular reach to which we can safely test axial symmetry with strong statistics.

\subsection{Data Acquisition}

The relevant data for this analysis was obtained from the European Space Agency's {\it Gaia} space telescope \citep{prusti2016gaia}, via the online archive\footnote{\url{https://gea.esac.esa.int/archive/}}.  Selections were made in windows of $(l, b)$ with $\varpi_{\rm unshifted} > -0.07$ mas and $|b| > 30^{\circ}$.  The latter cut is made in order to stay above the dust and gas close to the mid-plane while the former was made so that after a global shift of 0.07 mas is applied to the parallax, we have only stars with physical parallaxes, 
so that in the shifted parallax, $\varpi > 0$ mas.  We assume
this shift in all that follows. 
The particular value for the parallax zero-point shift was chosen from a wide array offered in the literature \citep{lindegren2018gaia, stassun2018evidence, zinn2019confirmation} such that our value is a ``middle-ground" choice.  The particular choice of shift 
has no bearing on the final results
as our data is quite close (the mean distance to a star in our sample is 0.94 kpc, with $\sim$60\% of the stars within 1 kpc), and thus does not bring previously negative parallaxes into the analysis.  Additionally, our analysis requires that a distance estimate via a parallax is available, so that we use only data for which the full five-parameter astrometric solution is available, as the two-parameter solution does not specify a distance \citep{luri2018gaia}.  After the analysis of the following two subsections, we conclude that elimination of the stars with the 2-parameter ``fallback" solution does not bias our result.

\subsection{Masking toward select anomalous lines of sight}

\begin{figure}[ht!]
\begin{center}
\subfloat[]{\includegraphics[scale=0.45]{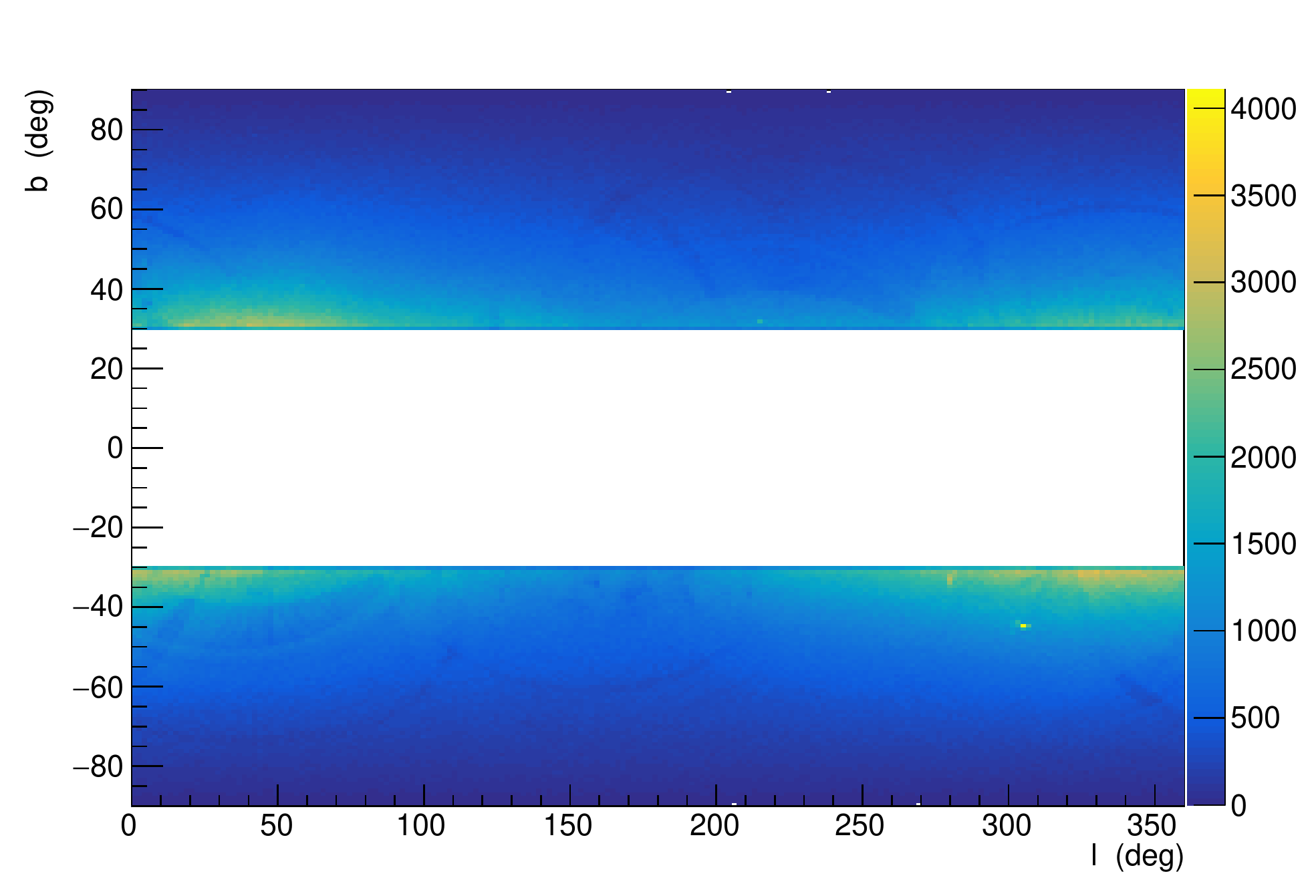}}
\subfloat[]{\includegraphics[scale=0.45]{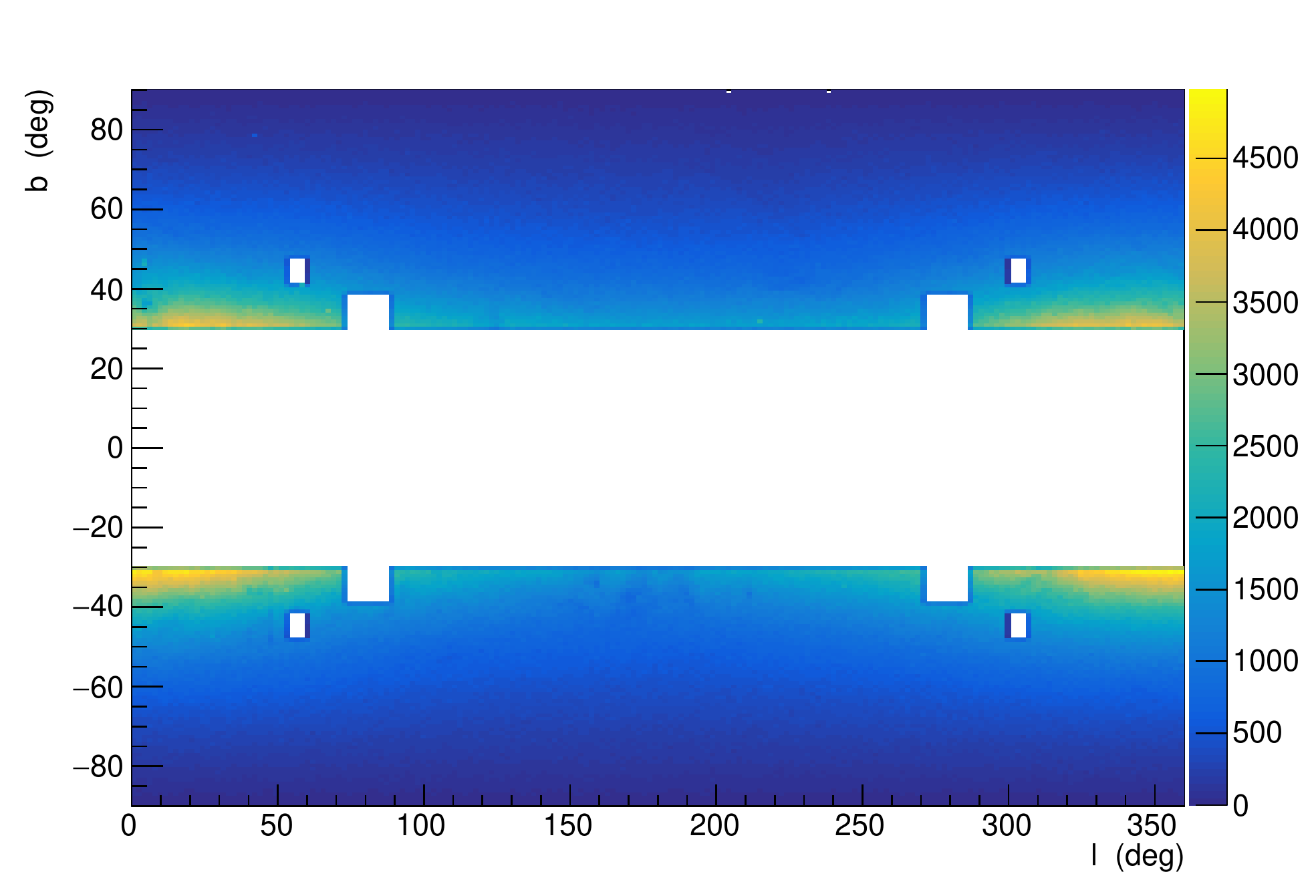}}
\caption{
(a) A selection of data in $(l,b)$ with 
relative parallax error cuts of $\sfrac{\sigma_{\varpi}}{\varpi} <$ 0.2 applied.  Notice the LMC and SMC ``bleed" through, even though their stars are at significantly greater distances than expected from a simple lower limit cut on the parallax with its error; stellar identification issues in the crowded field may be the cause of this issue.  There are also streaks of incompleteness in the data, a sign that {\it Gaia} did not measure stars in some regions of the sky with the same accuracy as others. 
(b) A selection of data in $(l,b)$ with the LMC and SMC removed, as well as all reflections of the two satellite galaxies to prevent any bias in an assessment of L/R and N/S symmetry.  
The cuts used are as follows:  $14 < G < 18$ mag, $0.5 < G_{BP} - G_{RP} < 2.5$, $\varpi > 0$ mas, $|b| > 30^{\circ}$, 0.2 $< |z| <$ 3 kpc, and the LMC/SMC excisions as outlined in Eqs.~\ref{LMCcuts} and \ref{SMCcuts}.
We note that some dust seems to survive the $|b|$ and $|z|$ cuts near the anti-center direction.  We have checked our final result with and without excisions of this region and there appears to be no appreciable effect.
}
\label{fig:lbcuts}
\end{center}
\end{figure}

In order to ensure our sample is not biased in the North vs. South or for $\phi > 180^{\circ}$ or $\phi < 180^{\circ}$, we are forced (see below) to cut out the Large and Small Magellanic Clouds, as well as their reflections across the mid-plane, the reflections across the anti-center line, and reflections across both the mid-plane and anti-center line, as in Fig.~\ref{fig:lbcuts} b).  We note here that we choose not to apply relative parallax error quality cuts as suggested by \citet{luri2018gaia}, as it is clear from Fig.~\ref{fig:lbcuts} a) that such cuts actually fail to remove the LMC and SMC from the data.  Additionally, visible streaks appear in the completeness data if we apply these parallax error cuts, preventing an unbiased assessment of axial symmetry.  

The excision of the LMC \& SMC was checked in a series of $(l,b)$ plots with the $G$, $G_{\rm BP} - G_{\rm RP}$, $|b|$, and $|z|$ restrictions outlined below already in place.  Thus, these selections act as 
an effective distance cut,
which helps with the elimination of the LMC/SMC contamination caused by poor parallax assessments in the dense field.  The LMC \& SMC sight lines, along with the appropriate reflections we have mentioned, are  removed by excising all the data that fit the following criteria:
\begin{equation}
    |b| \in [30^{\circ},39^{\circ}] \land \bigg(l \in [271^{\circ},287^{\circ}] \lor l \in [73^{\circ},89^{\circ}]\bigg)
    \label{LMCcuts}
\end{equation}
\begin{equation}
    |b| \in [41^{\circ},48^{\circ}] \land \bigg(l \in [299^{\circ},307^{\circ}] \lor l \in [53^{\circ},61^{\circ}]\bigg)
    \label{SMCcuts}
\end{equation}
for the LMC \& SMC, respectively. 

Although some dust seems to survive our $|b|$ and $|z|$ cuts in the region beyond the Sun's radius, we have checked that our asymmetry results do not change appreciably if that region is excised from our data set.

\begin{figure}[ht!]
\begin{center}
\subfloat[]{\includegraphics[scale=0.45]{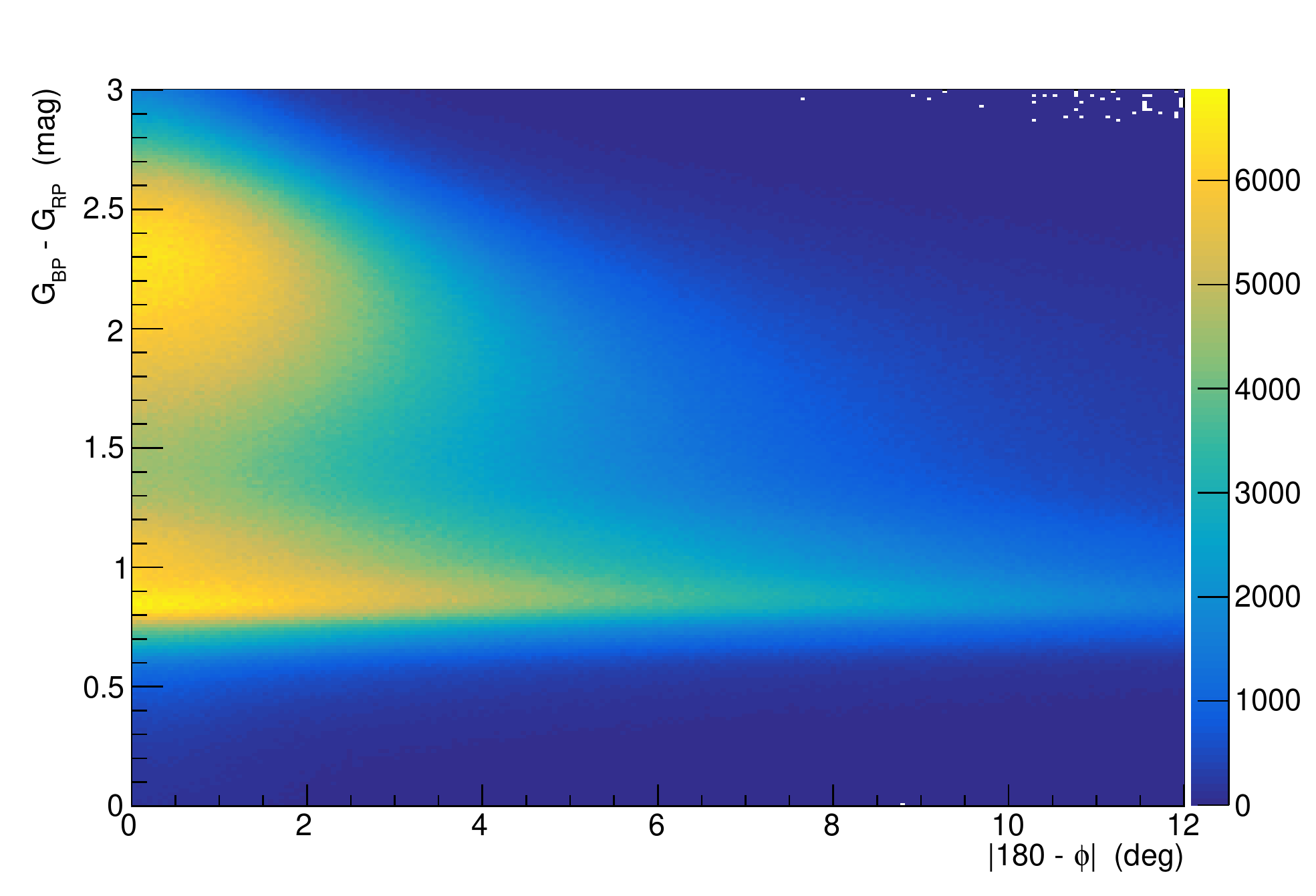}}
\subfloat[]{\includegraphics[scale=0.45]{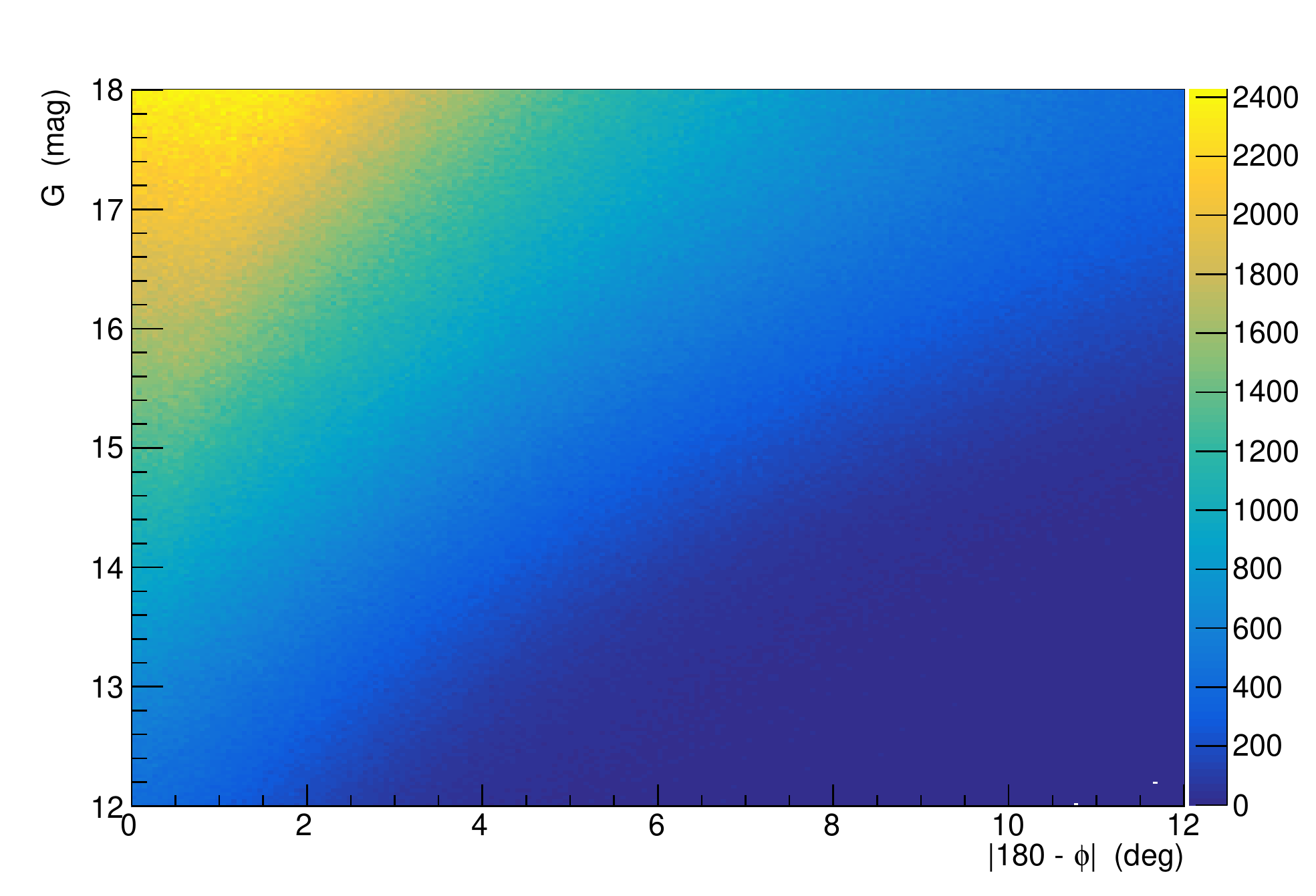}}
\caption{
(a) $G_{\rm BP} - G_{\rm RP}$  versus $|180^{\circ}-\phi|$.  For $0.5 < G_{\rm BP} - G_{\rm RP} < 2.5$ there appears to be significant statistical strength out to $|180^{\circ}-\phi| \approx 6^{\circ}$.  The cuts used are as follows:  $\varpi > 0$ mas, $|b| > 30^{\circ}$, and the LMC/SMC excision outlined in Eqs.~\ref{LMCcuts} and \ref{SMCcuts}. 
We note 
that the final results do not change appreciably upon changing the color cut from $G_{\rm BP} - G_{\rm RP} < 2.5$ mag to $G_{\rm BP} - G_{\rm RP} < 2.3$ mag.
(b) G band magnitude versus $|180^{\circ}-\phi|$.  For $G > 14$ mag there appears to be significant statistical strength out to $|180^{\circ}-\phi| \approx 6^{\circ}$. The cuts used are as follows: $0.5 < G_{\rm BP} - G_{\rm RP} < 2.5$ mag, $\varpi > 0$ mas, $|b| > 30^{\circ}$, and the LMC/SMC excision outlined in Eqs.~\ref{LMCcuts} and \ref{SMCcuts}.
}
\label{fig:colorAndMag}
\end{center}
\end{figure}

In order to assess the completeness of our sample after the implementation of the above ``box" cuts, the distribution of stars with respect to the angle $|180^{\circ} - \phi|$ was examined for multiple different parameters.  Beginning with the color and G-band magnitudes of our sample, as depicted in Fig.~\ref{fig:colorAndMag} a) and b) respectively, we see that there are various ranges of both parameters where we simply lack enough stars to probe very far from the $\phi = 180^{\circ}$ line.  In the interest of obtaining both a large number of stars and a non-negligible reach in $|180^{\circ} - \phi|$, we choose to keep stars for which $G_{\rm BP} - G_{\rm RP} \in [0.5,2.5]$ mag and $G \geq 14$ mag.  We also note that the distribution shows no signs of any obvious incompleteness.

\subsection{Limits on $G$ magnitude, $G_{\rm BP} - G_{\rm RP}$ color, $R$, and $|z|$}

\begin{figure}[ht!]
\begin{center}
\subfloat[]{\includegraphics[scale=0.45]{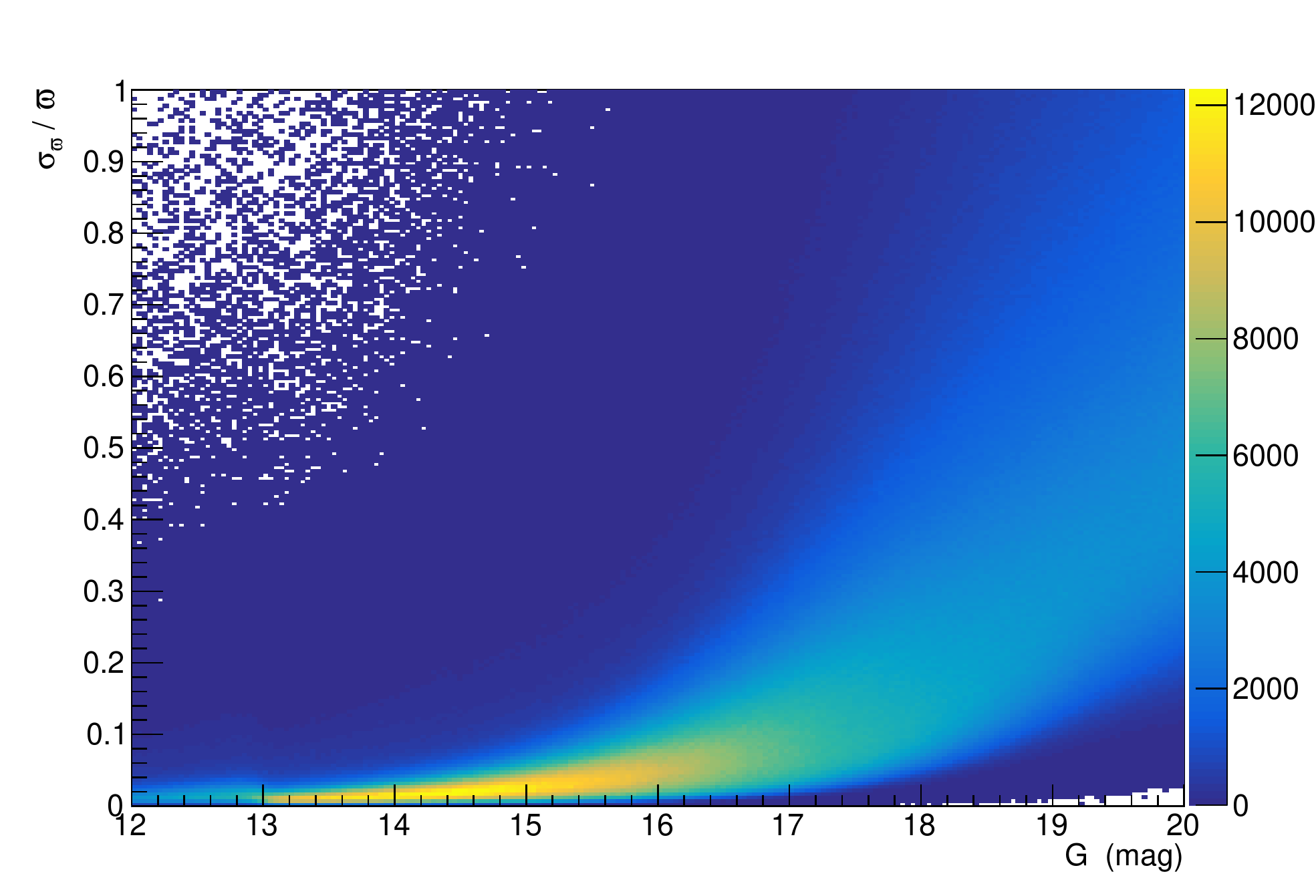}}
\subfloat[]{\includegraphics[scale=0.45]{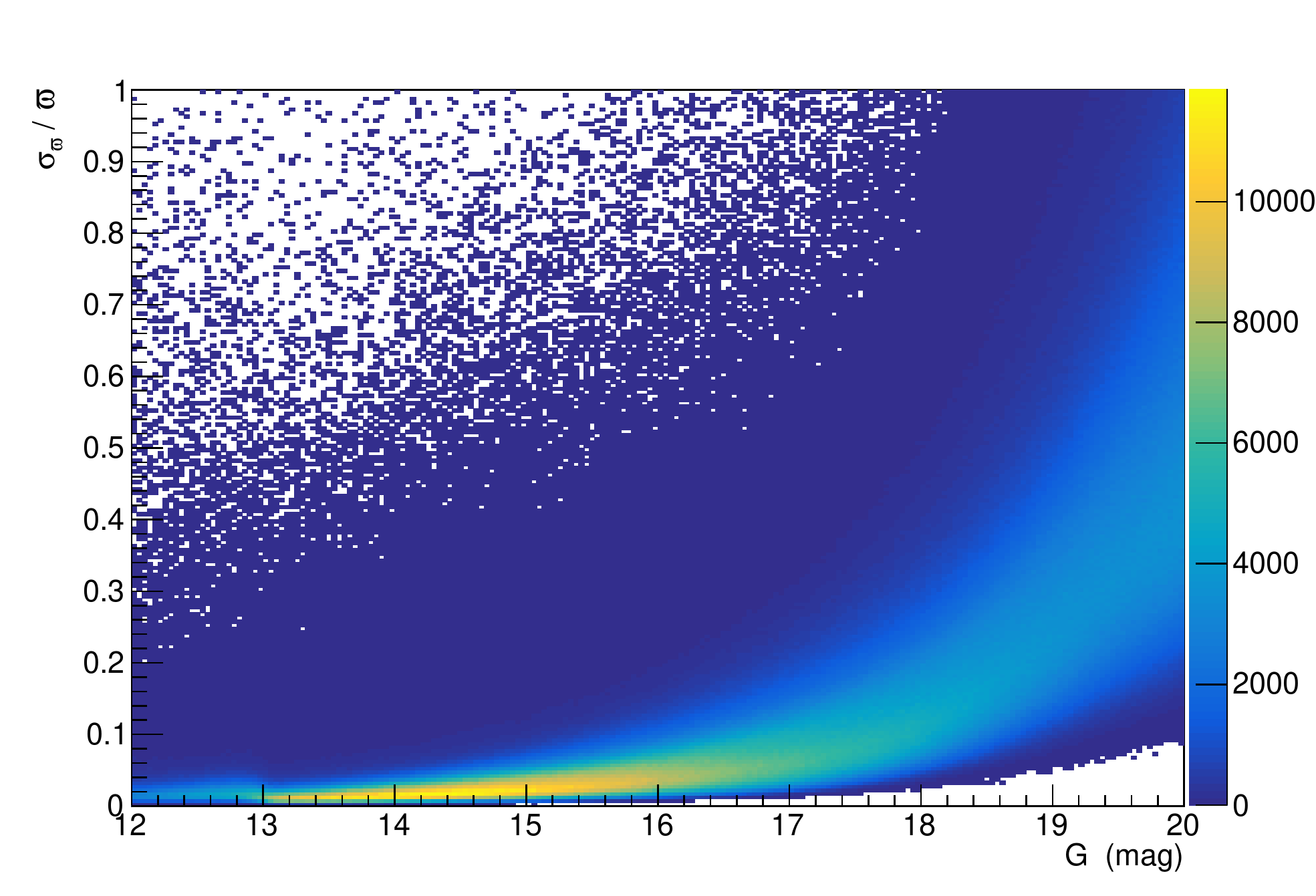}}
\caption{
(a) The majority of stars with magnitude $G$ have a relative parallax error that is reasonably small for $G < 18$ mag.  The {\it Gaia} documentation mentions incompleteness in crowded fields and due to data processing and  ``filtering" for stars with $G > 17$ mag in crowded regions \citep{arenou2018gaia}; however, we avoid these regions and can thus extend our reach to fainter stars, as motivated in the text.
The cuts used are as follows: $0.5 < G_{\rm BP} - G_{\rm RP} < 2.5$ mag, $\varpi > 0$ mas, $|b| > 30^{\circ}$, and the LMC/SMC excision outlined in Eqs.~\ref{LMCcuts} and \ref{SMCcuts}. These are the 
``standard cuts'' that we employ throughout our analysis. 
(b) The relative error in the parallax for our selection of stars, though without $G$ restrictions.  We manage to select stars with relatively well measured parallaxes without incurring the bias of cutting on relative parallax error explicitly. The cuts used are as follows: $0.5 < G_{\rm BP} - G_{\rm RP} < 2.5$ mag, $\varpi > 0$ mas, $|b| > 30^{\circ}$, $7 < R < 9$ kpc, and the LMC/SMC excision outlined in Eqs.~\ref{LMCcuts} and \ref{SMCcuts}.
}
\label{fig:Gprlxerr}
\end{center}
\end{figure}

Although Fig.~\ref{fig:colorAndMag}b) suggests that one may probe larger values of $|180^{\circ}-\phi|$ for fainter stars, one runs into unreliable parallax measurements for $G > 18$ mag, as the majority of stars in the trend from Fig.~\ref{fig:Gprlxerr} with $G > 18$ mag have $\sfrac{\sigma_{\varpi}}{\varpi} > $ 0.2.  As such, we consider only the window $G \in [14, 18] $ mag.  While similar to the suggestion of \citet{luri2018gaia} that relative parallax error cuts be used, we instead rely on the magnitude as a rough proxy for the relative 
parallax error.
Because we are looking away from the gas and dust of the plane due to our requirement that $|b| > 30^{\circ}$, we expect that the magnitudes observed by {\it Gaia} will not dramatically fluctuate over small scales like a hard cut on $\sfrac{\sigma_{\varpi}}{\varpi}$ did in Fig.~\ref{fig:lbcuts} a). 
We are mindful that \citet{luri2018gaia} suggest cutting out all stars with $G > 17$ mag due to incompleteness in crowded fields, the scan-law pattern, and filtering during the data reduction process.  We show later in Fig.~\ref{fig:Aggregate} that the quantitative result found for the smaller, recommended magnitude window is not very different than the $G \in [14,18]$ mag window result. Since we have cut out the regions with crowding issues (i.e., low latitudes and the LMC and SMC), using $G<18$ mag appears to be a valid option for our particular set of cuts.  By using this faintness limit, we are adding statistical strength to our analysis.

\begin{figure}
    \centering
    \includegraphics[scale=0.5]{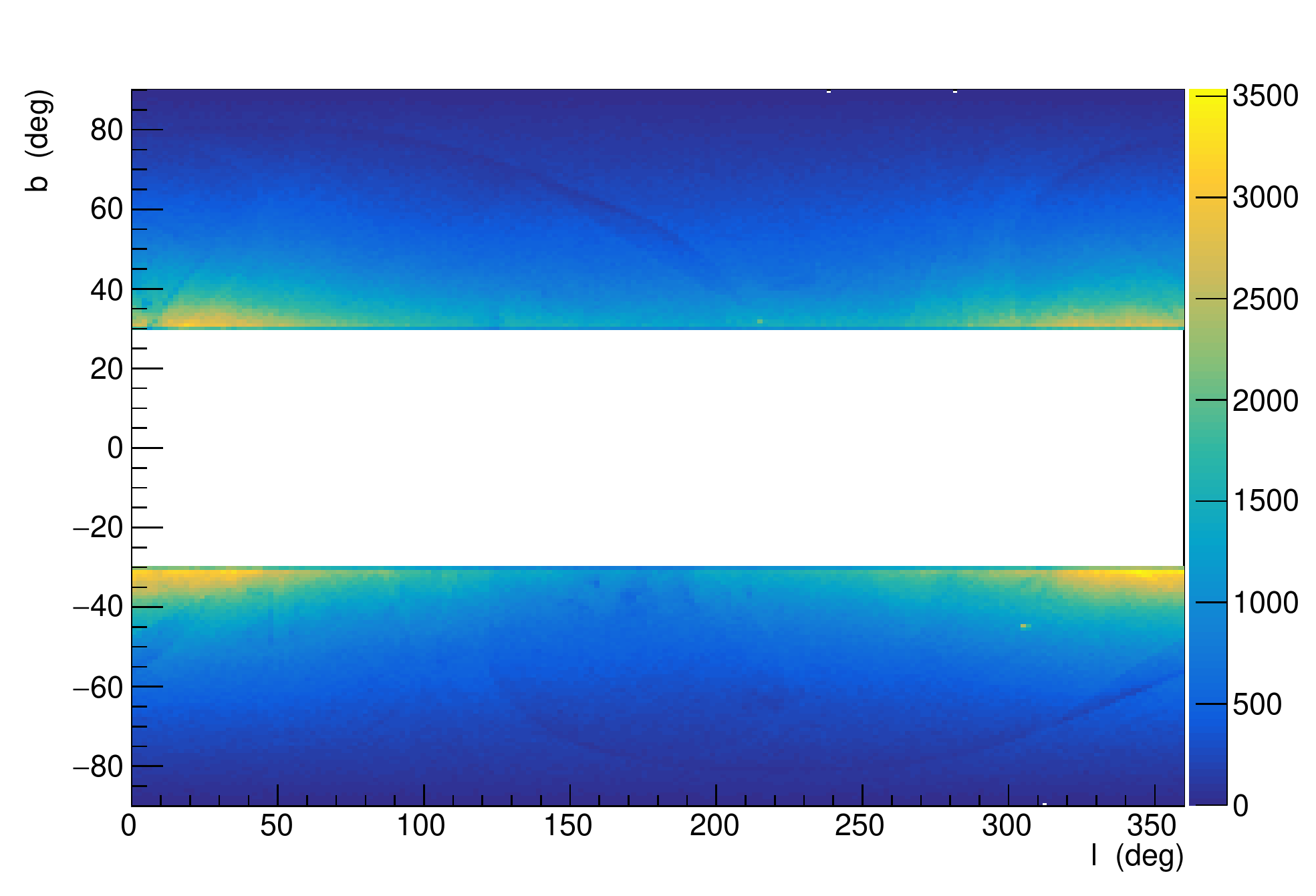}
    \caption{A study of stars within $\lesssim 3$ kpc with $|b|>30^{\circ}$ and $G \in [14,18]$ mag,
    but with restrictions on Astrometric Excess Noise (AEN) in place.  All stars with an AEN value larger than 0.2 mas have been excised.  Clearly, requiring such a quality cut would leave the sample 
    with artificial asymmetries 
    due to the streaks seen in many portions of the sky.}
    \label{fig:AENcuts}
\end{figure}

One possible alternative is using the Astrometric Excess Noise (AEN),  available in the {\it Gaia} data \citep{lindegren2012astrometric}, rather than the relative error in parallax, as it is not explicitly a function of the number of observations \citep{lindegren2018gaia},
which are known to be different in different parts of the sky \citep{lindegren2018gaia}.  However, as Fig.~\ref{fig:AENcuts} illustrates, there is a significant bias incurred from restricting the values of the AEN.  As such, we forgo any restrictions on the AEN, and instead check that our final data set results in a 
relatively small 
average AEN.  For our particular latitudes, the mean AEN, $\langle{\rm AEN} \rangle$, is about 1.034 mas for the entire {\it Gaia} database, whereas a similar yet slightly larger volume of space to our analysis (i.e., $\varpi > 0.3$ mas and $|b|>30^{\circ}$) with stars having similar magnitudes and colors yields an upper bound value of $\langle{\rm AEN} \rangle \approx$ 0.141 mas, nearly an order of magnitude better.  

\begin{figure}[ht!]
\begin{center}
\subfloat[]{\includegraphics[scale=0.45]{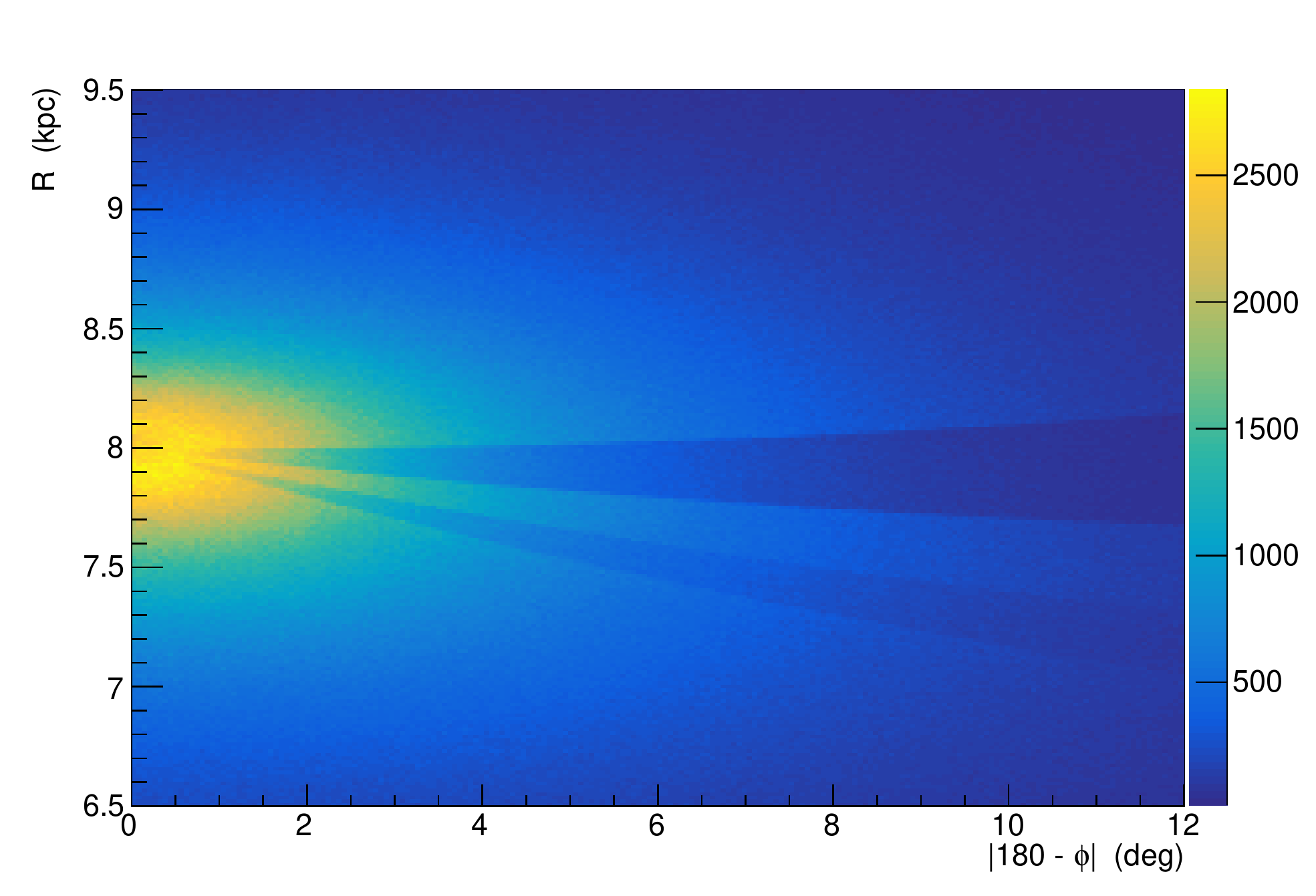}}
\subfloat[]{\includegraphics[scale=0.45]{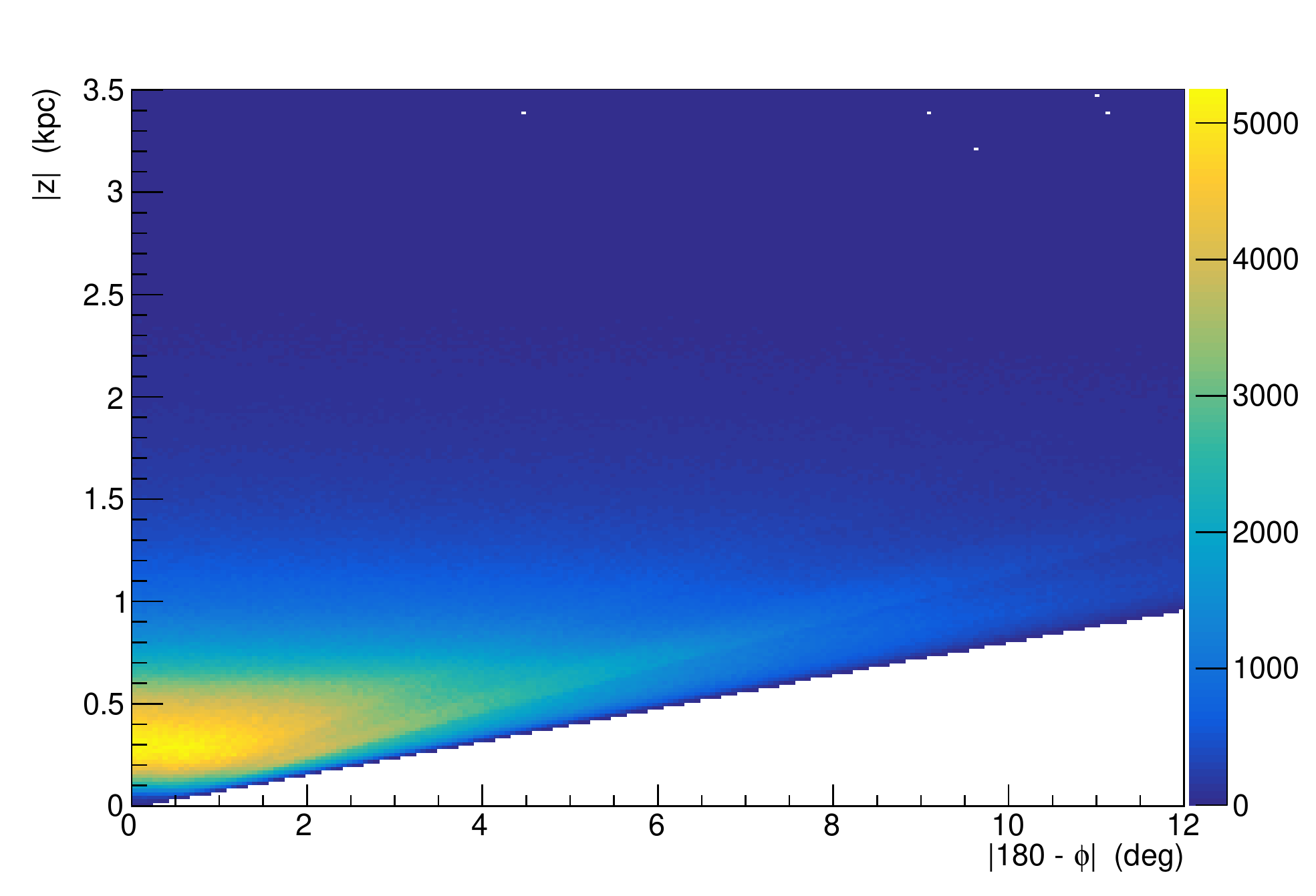}}
\caption{
(a) $R, \phi$ completeness for the selected data. We choose to cut on $7 < R < 9$ kpc in order to achieve the best angular reach possible without compromising completeness. The cuts used are as follows: $14 < G < 18$ mag, $0.5 < G_{\rm BP} - G_{\rm RP} < 2.5$ mag, $\varpi > 0$ mas, $|b| > 30^{\circ}$, and the LMC/SMC excision outlined in Eqs.~\ref{LMCcuts} and \ref{SMCcuts}.
(b) Test of vertical completeness over $\phi$.  Cuts used: $14 < G < 18$ mag, $1.5 < G_{\rm BP} - G_{\rm RP} < 2.5$ mag, $7 < R < 9$ kpc, $\varpi > 0$ mas, and the LMC/SMC excision outlined in Eqs.~\ref{LMCcuts} and \ref{SMCcuts}.
}
\label{fig:RandZcompleteness}
\end{center}
\end{figure}

Additionally, the distribution of our data set is checked over a range of in-plane radius, $R$, in Fig.~\ref{fig:RandZcompleteness} a).  From this examination, we cut on $R \in [7, 9]$ kpc in order to circumvent sampling 
a region of low statistical strength.  To avoid the stars too close to the plane with known axially asymmetric features (e.g. spiral arms), we make a cut of $|z| > 0.2$ kpc.  Also, to avoid straying too far from the disk into incomplete regions of $|z|$, we require $|z| < 3$ kpc as motivated by Fig.~\ref{fig:RandZcompleteness} b), where we admittedly push the data to its limits in order to sample more halo stars.

\begin{figure}[ht!]
\begin{center}
\subfloat[]{\includegraphics[scale=0.45]{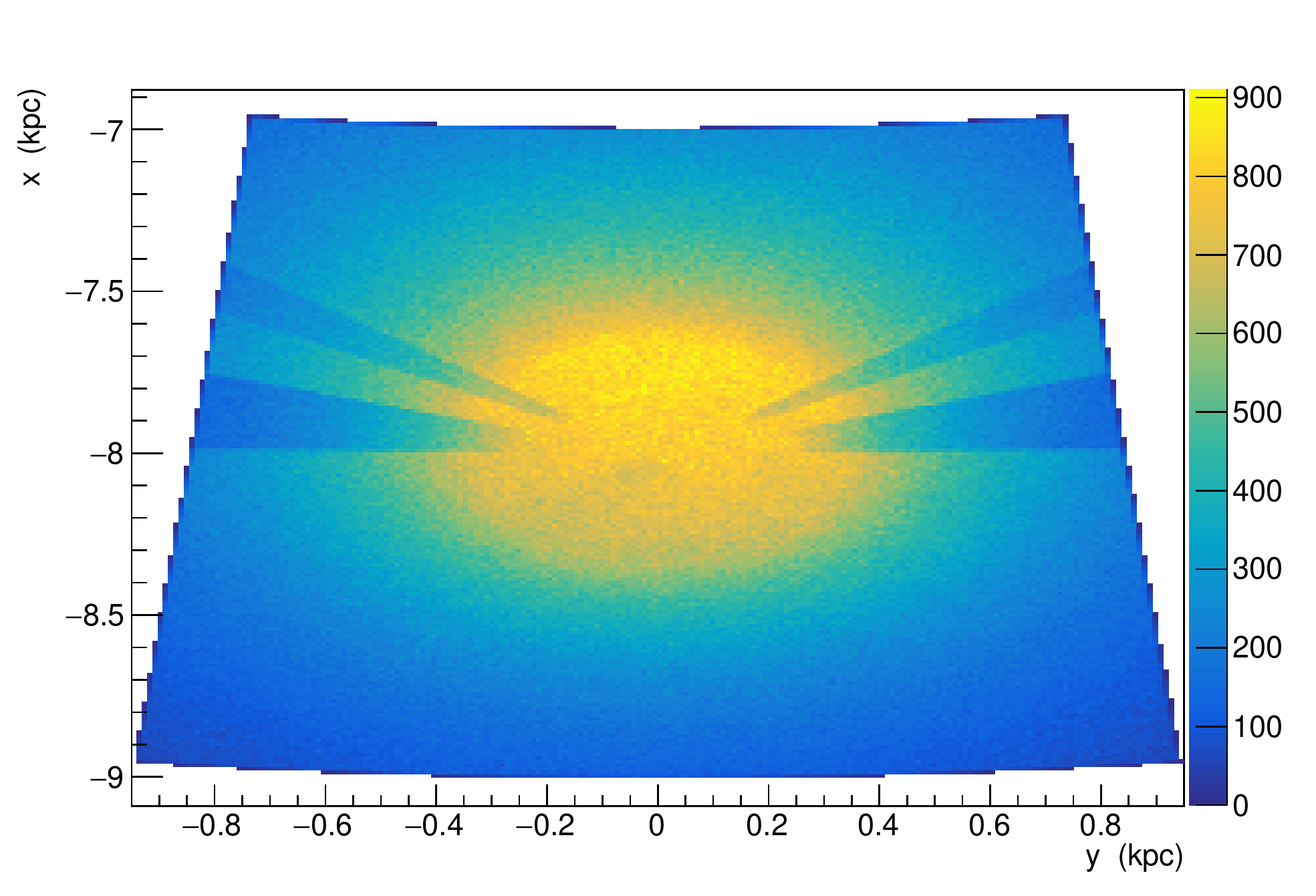}}
\subfloat[]{\includegraphics[scale=0.45]{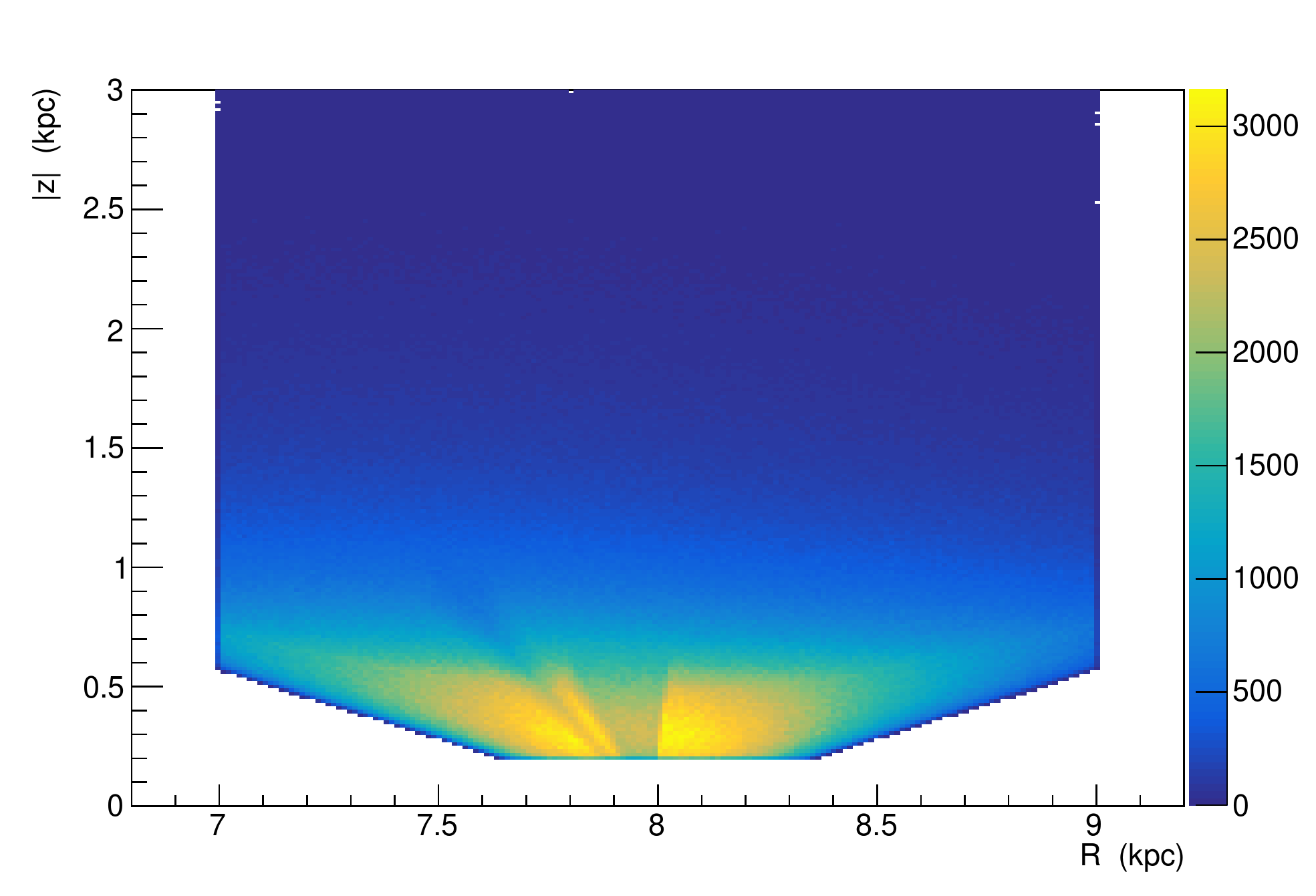}}

\subfloat[]{\includegraphics[scale=0.45]{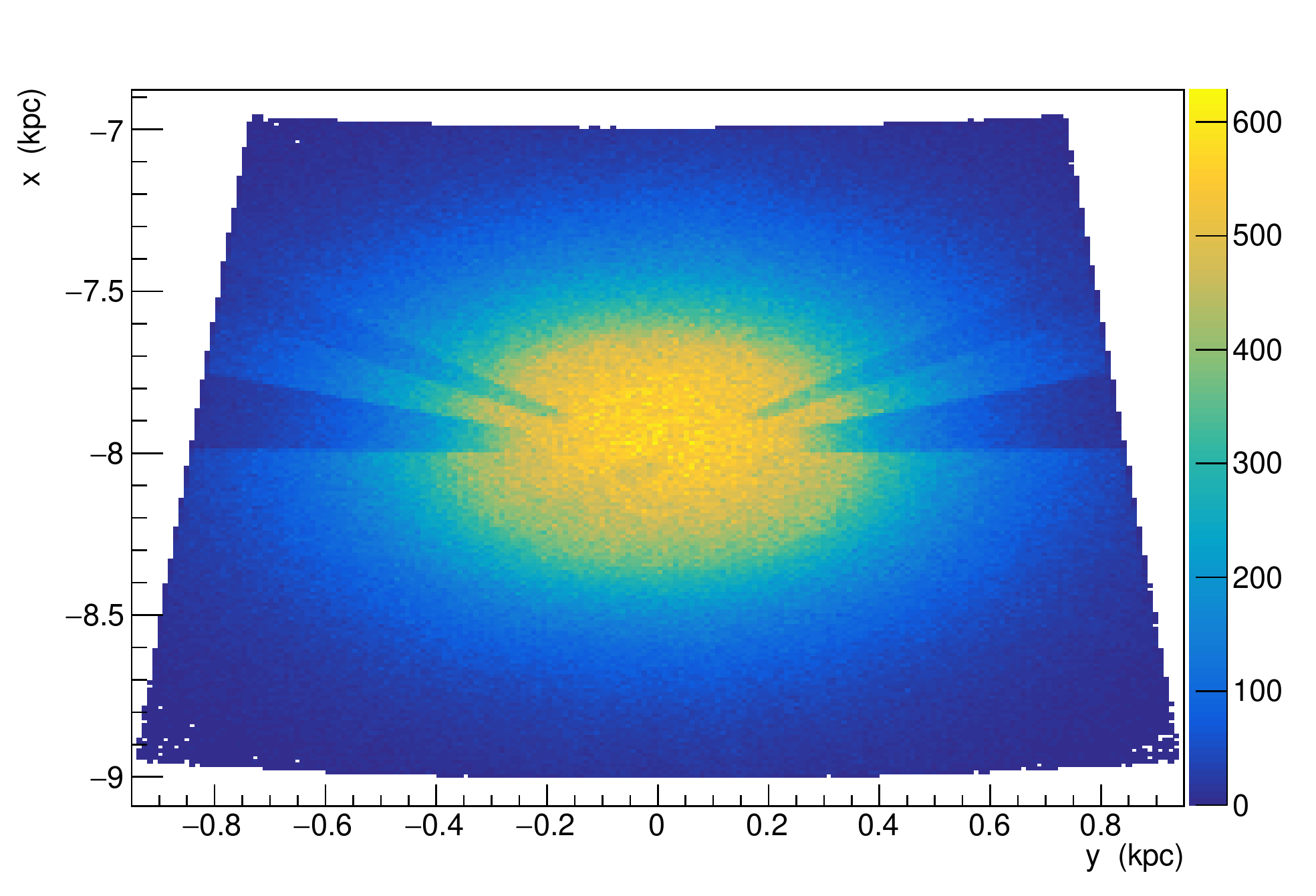}}
\subfloat[]{\includegraphics[scale=0.45]{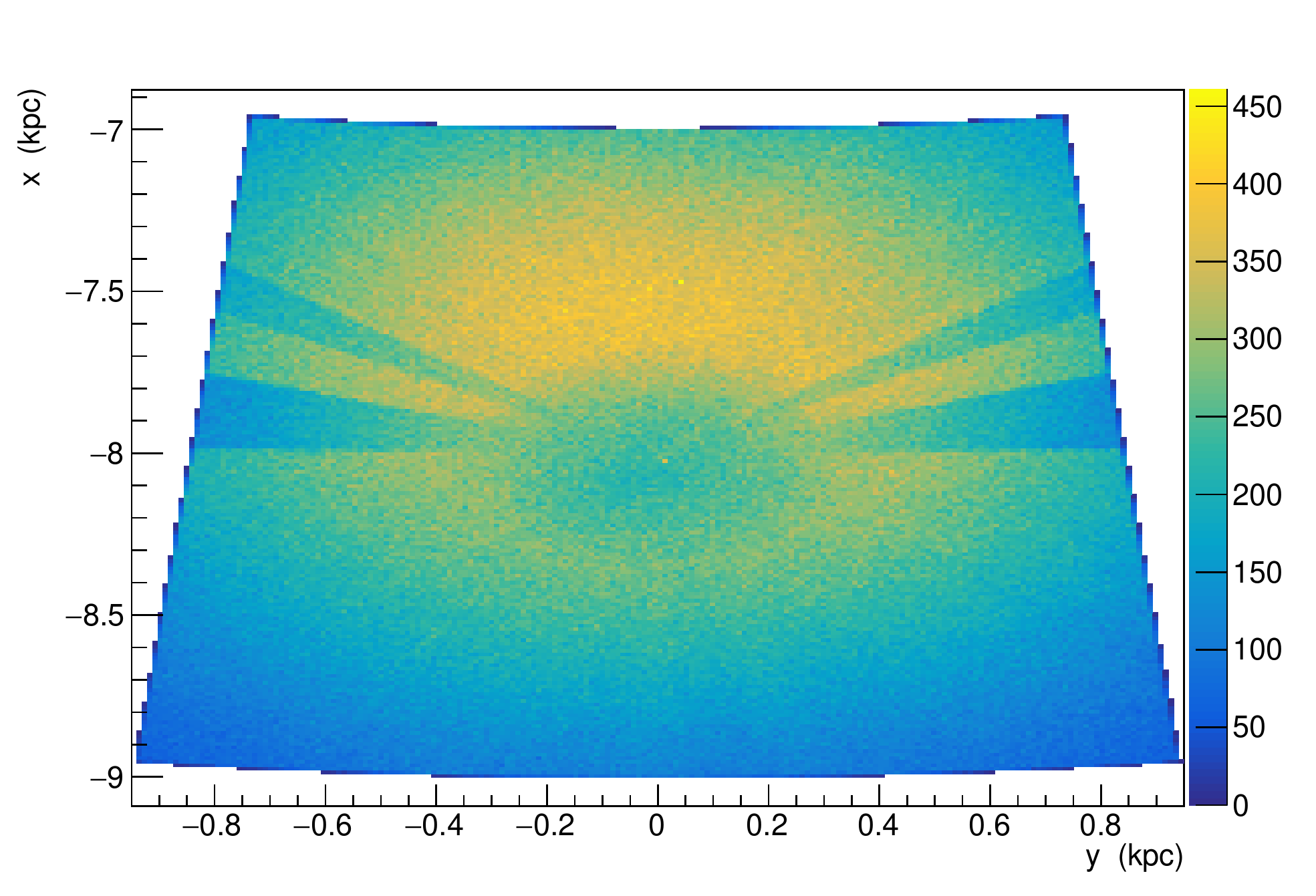}}
\caption{
(a) The in-plane projection of our data with the cuts motivated above, as viewed from the South side of the plane.  The data extends out to 6 degrees in galactocentric azimuth.  
(b) The geometry of our data selection in the $R-|z|$ plane. 
(c) The in-plane projection for red stars ($1.5 < G_{\rm BP} - G_{\rm RP} < 2.5$ mag).
(d) The in-plane projection for blue stars ($0.5 < G_{\rm BP} - G_{\rm RP} < 1.5$ mag).
All panels have the following cuts unless noted otherwise:  $0.5 < G_{\rm BP} - G_{\rm RP} < 2.5$ mag, $14 < G < 18$ mag, $\varpi > 0$ mas, $|b| > 30^{\circ}$, $7 < R < 9$ kpc, $0.2 < |z| < 3.0$ kpc, $|180^{\circ} - \phi| \leq 6^{\circ}$, and the LMC/SMC excision outlined in Eqs.~\ref{LMCcuts} and \ref{SMCcuts}.
}
\label{fig:geometry}
\end{center}
\end{figure}

With all of the above constraints applied, we are left with a data set which is free of artificial biases
left and right of the $\phi = 180^{\circ}$ line.  Not only is the data set well-matched, the stars themselves are, on average, very well measured with a mean relative uncertainty in parallax of about 8.6\% (see Table~\ref{tab:relativeerrors} for relative uncertainty in parallax for each quadrant).  In order to visualize this resulting
data set, we show the in-plane projection of our data in Fig.~\ref{fig:geometry} a).  The geometry of the data in the $R-|z|$ plane, on the other hand, is depicted in Fig.~\ref{fig:geometry} b).  Figs.~\ref{fig:geometry} c) and d) show the in-plane reach for red and blue stars respectively, where one can see that the maximal reach in $|180^{\circ} - \phi|$ is reduced slightly for the redder population.

\begin{center}
\begin{table}[ht!]
    \centering
    \caption{\label{tab:relativeerrors} The relative uncertainty in parallax, $\sfrac{\sigma_{\varpi}}{\varpi}$, for each quadrant of the analysis after applying the standard set of cuts enumerated in 
    Fig.~\ref{fig:Gprlxerr} with $R\in [7,9]$ kpc and 
    $z \in [0.2, 3]$ kpc --- and as also employed in \citet{GHY20}.
    }
    \hskip-2.45cm  
    \begin{tabular}{ |c|c|c|c| } 
       \hline
       \ & Left & Right & Total\\ 
       \hline
       North & 0.101 & 0.074 & 0.087\\ 
       \hline
       South & 0.072 & 0.098 & 0.085\\ 
       \hline
       Total & 0.086 & 0.086 & 0.086\\ 
       \hline
   \end{tabular}
\end{table}
\end{center}

\subsection{Consideration of Systematics}

In order to assess the impact of various systematic effects on our data, we explore a number of different possibilities and estimate the degree to which they could cause a false asymmetry signal.  First, due to the fact that we use Galactic azimuth values for each star, this necessitates that a parallax and thus a distance is known for the star.  As such, this requirement eliminates all sources in the {\it Gaia} database which have a 2D ``fall-back" astrometric solution only.  Further, as we require color cuts for the completeness arguments we have outlined, we also miss stars with null entries for $G_{\rm BP} - G_{\rm RP}$.  We have explicitly checked via a statistical query of the {\it Gaia} database that the asymmetry such an effect could cause would be $|{\cal A}_{\rm null \  \varpi \ \& \ color}| \approx 0.0017$ for our $G$ and $|b|$ requirements.
However, 
the preponderance of these stars lie outside our sample volume.
Indeed, in a Bayesian analysis to 
associate a distance with the 2D fallback solution, a prior used for the stars with poor astrometric measurements assumes a small parallax \citep{michalik2015gaia}, and therefore a large distance. 
Moreover, as we discuss further below, a Hubble Space Telescope comparison study suggests 
our sample is essentially 100\% complete.  
Therefore, the value given above is a gross overestimate, 
and taking the estimated 
incompleteness conservatively into account suggests that 
the estimated size of this systematic effect to be no more 
than $|{\cal A}| \sim 2 \times 10^{-4}$.

Next, it is a well known feature of the {\it Gaia} telescope that its orbital motion corresponds to certain parts of the sky being imaged more than others \citep{lindegren2018gaia}, hence the streaks seen when relative parallax error restrictions are made \citep{luri2018gaia}.  However, this discrepancy in the number of observations does not affect whether or not a relatively close star is seen at all.  In fact, the completeness of the {\it Gaia} data for relatively close stars has been checked against 2MASS and has been found to be about 99\% complete for 
data selections 
not too dissimilar from ours \citep{bennett2018vertical}.  

Further, the completeness of the {\it Gaia} data has been checked against the Hubble Space Telescope for dense fields in \citet{arenou2018gaia} and, for the range of magnitudes and latitudes considered here, is essentially 100\% complete.  In other words, by cutting away the densest regions of the sky and the dimmest stars, we are not affected by the completeness issues that often plague stars with such parameters.  

Although the 
average angular 
densities considered in our analysis are all well within the range of safe values suggested by \citet{arenou2018gaia}, one might worry that 
very localized lines of sight could exceed these density limits and result in small incomplete regions.  We explicitly check this by zooming into the distribution of stars in $l,b$ to search 
for small over-densities. We do find such regions, but we estimate 
that the impact of the stars missed in such cases on ${\cal A}$ to be  extremely small.
We study a 
worst case scenario by studying stars near $|b|=30^\circ$ near 
the GC,
gauging the expected level of completeness as detailed by \citet{arenou2018gaia}.  We find very localized areas with source densities near $4 \times 10^5$ stars per square degree. 
Such a missing population of stars cannot possibly account for the symmetry breaking we see, as 
simply counting the overdense regions on each side of the $\phi = 180^{\circ}$ line yields an estimated
asymmetry of $|{\cal A}| \approx 0.000013$.  

Another possible systematic comes from the parallax offset.  To first order, the parallax offset is assumed to be the same for stars in all directions, magnitudes, colors, etc.  As such, there would be no differences left or right, and thus no contribution to the asymmetry.  If instead the shift varies slightly on the sky, a small, ${\cal O}(0.01)$ mas correction to nearby stars would correspond to a distance shift of about 10 pc, or about 1\% of the typical distance to a star in our sample.  If this effect were truly an issue, the completeness checks we have performed would likely have exhibited some sort of noticeable fluctuation as stars with a different, true offset would have shifted into or out of our sample; we see no evidence of this.

Finally, the {\it Gaia} data obviously exhibits scan lines when relative parallax uncertainty quality cuts are implemented, so that one may be tempted to 
think that scan lines may also be hidden in the completeness plots 
at a level not detectable to the human eye.  To test this possibility, a number of tests were devised.  First, upon zooming into small regions of $l,b$ in our data where there are known streaks of incompleteness in the separate, relative parallax error data, as in Fig.~\ref{fig:lbcuts} a), we see no signs of any scan lines.  Further, we numerically test this possibility by examining the expected asymmetry in our data set along the lines of sight toward a known streak in the relative parallax error data.  The local, worst-case asymmetry is $|{\cal A}| = 0.007$, and since the ``would-be" streaks constitute a small fraction of the sky and are on both sides of the $\phi = 180^{\circ}$ line, the expected, total asymmetry from any such effect is smaller still, perhaps something like a tenth of the worst case local measurement, $|{\cal A}| = 0.0007$.  
 
Tallying the impact of the various systematic effects we have considered
we estimate the total systematic error to be no larger than 
$|{\cal A}_{\rm sys}|\sim 0.0009$. We note that this is substantially 
smaller than the asymmetry effects we have observed.

\subsection{Splitting the sample into three radial bins}

\begin{figure}[ht!]
\begin{center}
\subfloat[]{\includegraphics[scale=0.55]{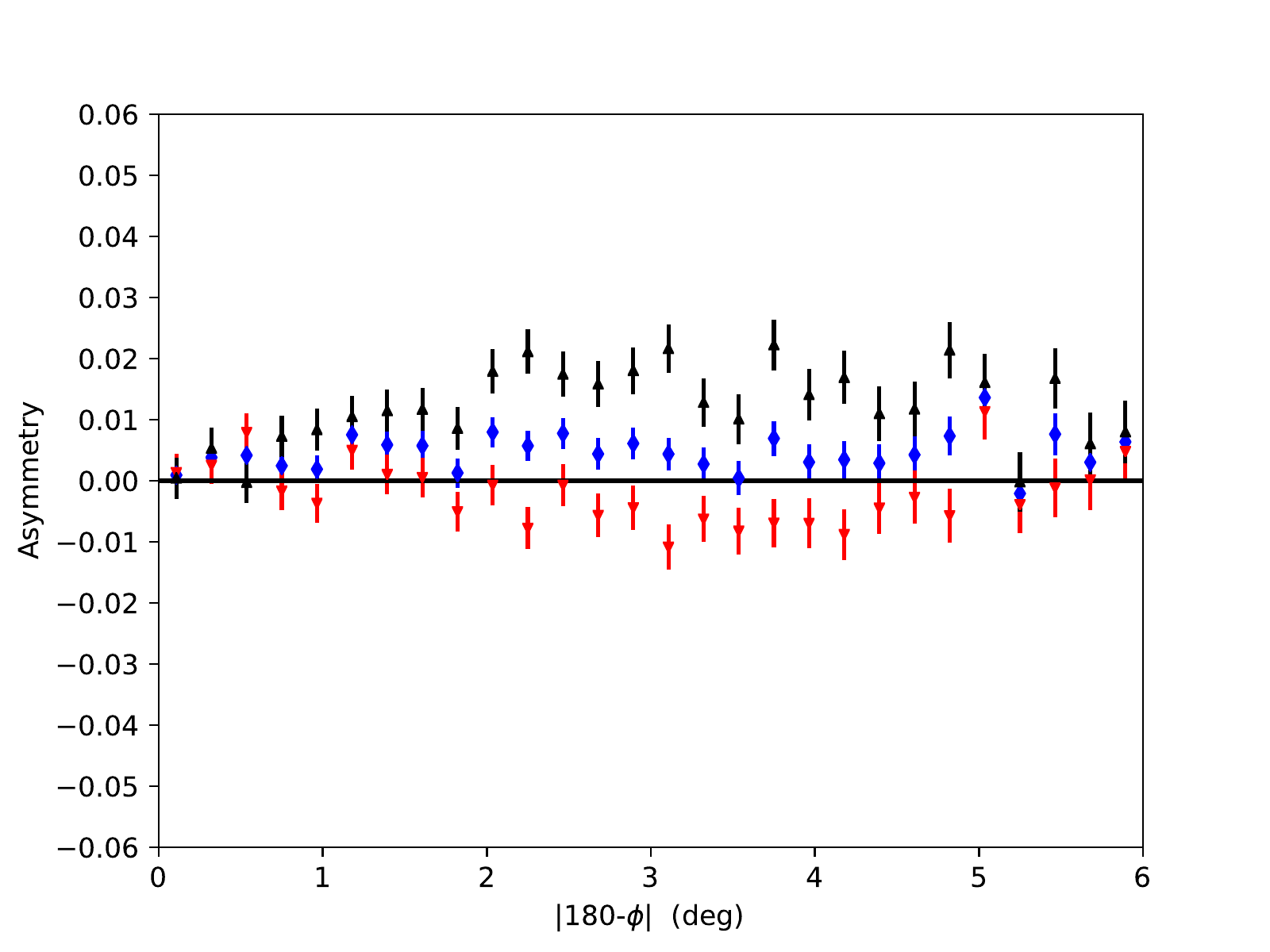}}
\subfloat[]{\includegraphics[scale=0.55]{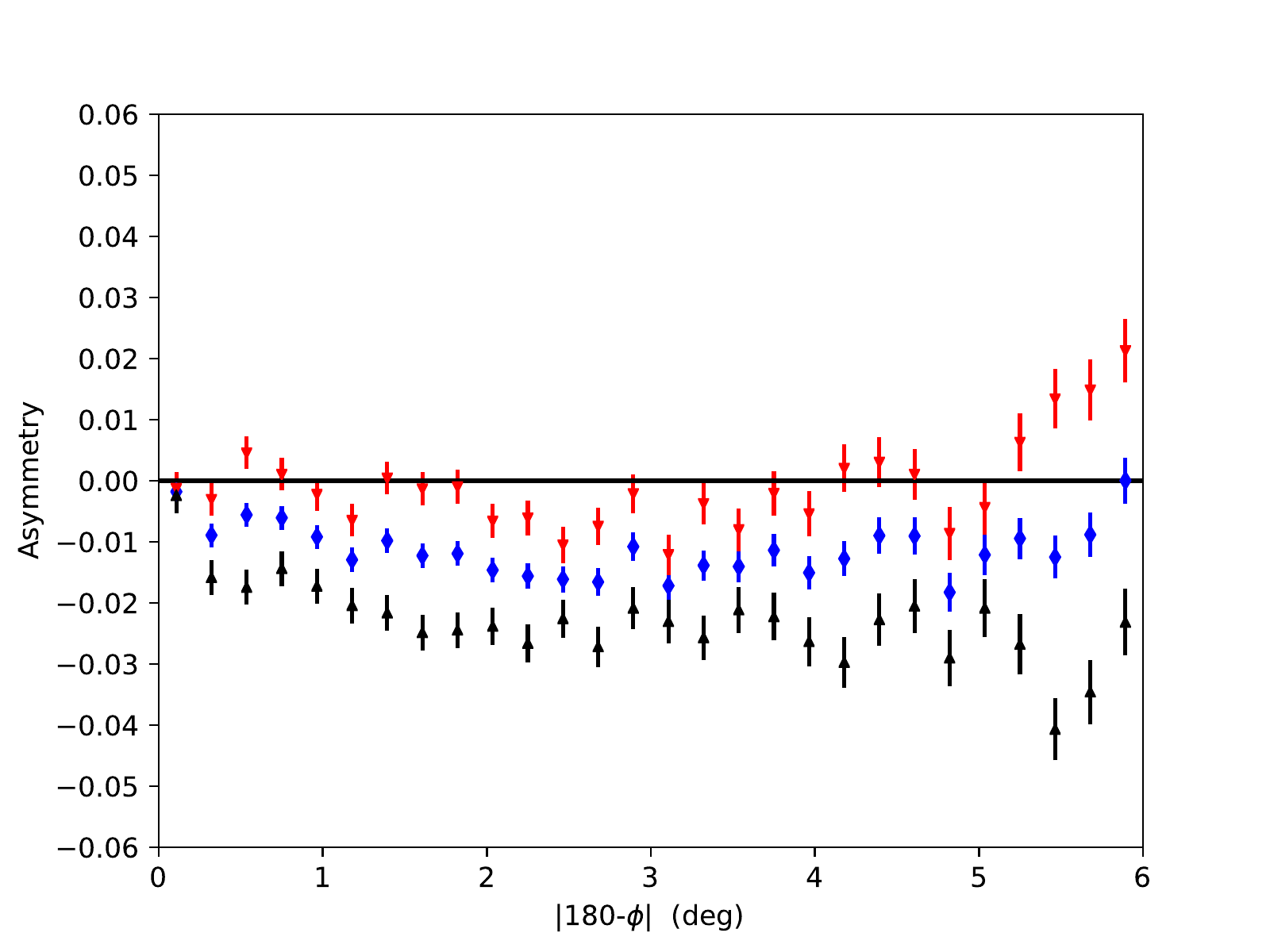}}

\subfloat[]{\includegraphics[scale=0.55]{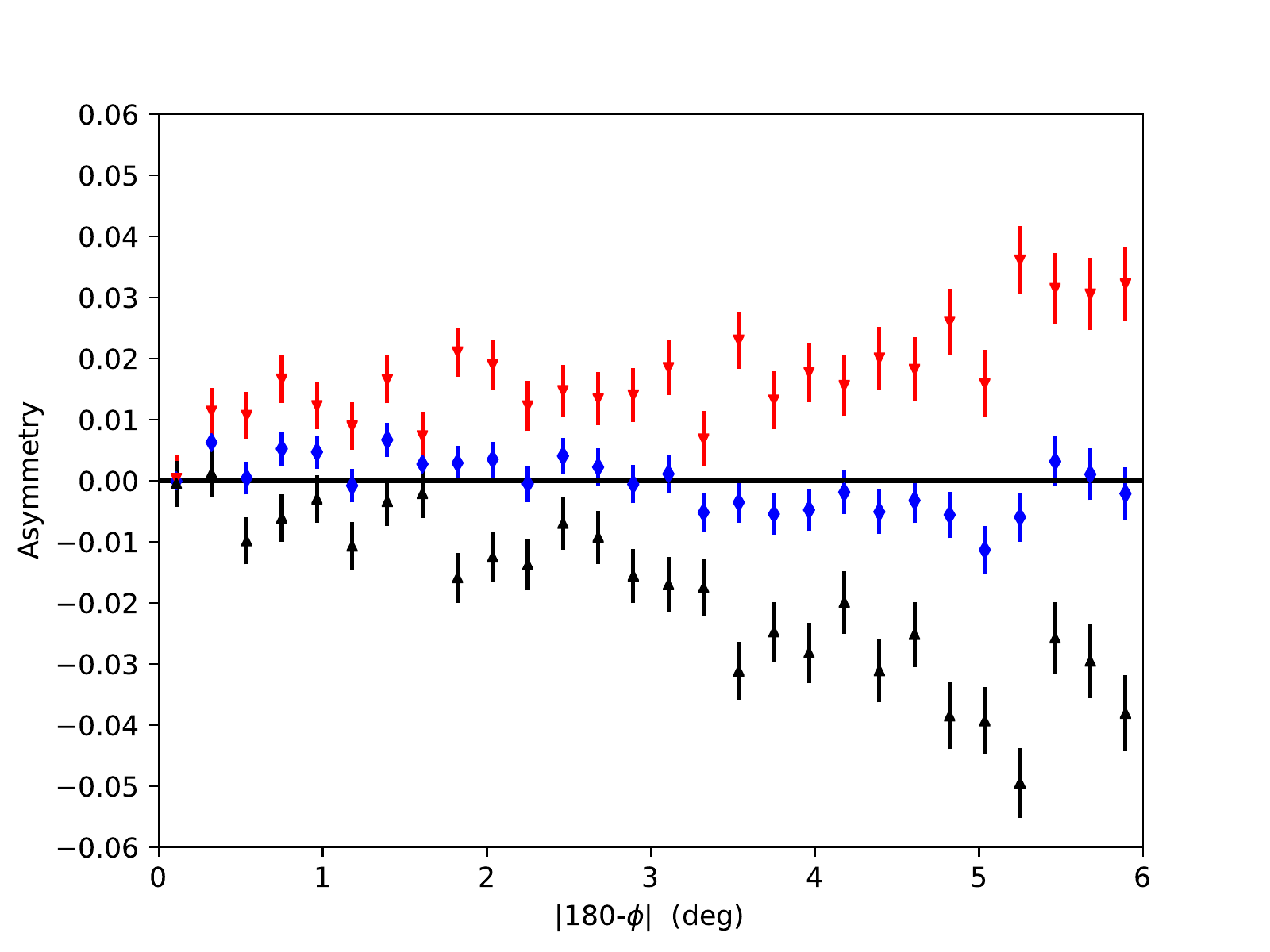}}
\caption{
Dividing the data set into radial bins unveils a radial dependence for the North + South (blue diamonds), North only (black upward triangles), and South only (red downward triangles).
(a)  The axial symmetry test for stars with $R \in [7,7.7]$ kpc.  
(b)  The axial symmetry test for stars with $R \in [7.7,8.3]$ kpc.  
(c)  The axial symmetry test for stars with $R \in [8.3,9]$ kpc.  
Additional cuts used in all panels are: $0.5 < G_{\rm BP} - G_{\rm RP} < 2.5$ mag, $14 < G < 18$ mag, $\varpi > 0$ mas, $|b| > 30^{\circ}$, $0.2 < |z| < 3.0$ kpc, $|180^{\circ} - \phi| \leq 6^{\circ}$, and the LMC/SMC excision outlined in Eqs.~\ref{LMCcuts} and \ref{SMCcuts}.
}
\label{fig:Rdepend_normal}
\end{center}
\end{figure}

Since \citet{GHY20} studied the axial asymmetries for our 
aggregate sample and argued for a dominant role by the LMC \& SMC
in interpreting the results seen, 
we first split the sample into three $R$ bins, $7 < R < 7.7$, $7.7 < R < 8.3$, and $8.3 < R < 9.0$ kpc in order to see if the asymmetry is constant with $R$.  
If the asymmetry were due entirely to the LMC, which because of its distance, would act over a relatively large coherent area on mass in the Milky Way, we would expect the ${\cal A}$ trends with $\phi$ to be similar in the three bins.
Perhaps surprisingly,
after dividing the sample into three radial bins, one instead notices marked differences in the shapes 
of the various asymmetry curves in Fig.~\ref{fig:Rdepend_normal}
with $\phi$.  These marked differences can potentially be understood as a superposition of two effects.  First, given that the disk scale length is $R_{s} \approx$ 2 kpc \citep{bovy2013direct}, the 
contribution of the halo to the total, N+S asymmetry (blue diamonds) is less pronounced in the inner region with more disk stars.  The second effect we resolve is in only the two lowest $R$ bins.  In the innermost case, the left-heavy asymmetry differs in sign 
from the model matter distribution with a distorted halo\citep{erkal2019total} that can confront the shape of 
the aggregate data fairly well~\citep{GHY20}.  Interestingly, such 
a $R$ variation in the asymmetry could qualitatively be an expected signal from an Outer Lindblad Resonance (OLR) of the Galactic bar, where we note Figure 1 of \citet{dehnen2000effect} and the explanations of \citet{contopoulos1980orbits}, or perhaps the Co-rotation resonance (CR) of the Galactic bar --- as we discuss further below.  Given that such effects are not axially symmetric, our method is sensitive to them and thus we may be able to discriminate between these two possibilities.

\subsection{Masking out the GC
and Removing Dimmer Stars} \label{subsec:GCcuts}

Given that the GC
direction has an extremely high number of stars per solid angle, the {\it Gaia} telescope may have issues accurately measuring parallaxes and correctly identifying stars in this region.  With this thought in mind, we check our radially-separated results by cutting away the stars most likely to be biased by such an effect.  Namely, we do this via three alternative methods:  lowering our faintness cut to $G < 17$ mag, excising the GC
via ``box" cuts in $l$ and $b$, and utilizing both of the aforementioned cuts in tandem. 

\begin{figure}[ht!]
\begin{center}
\subfloat[]{\includegraphics[scale=0.55]{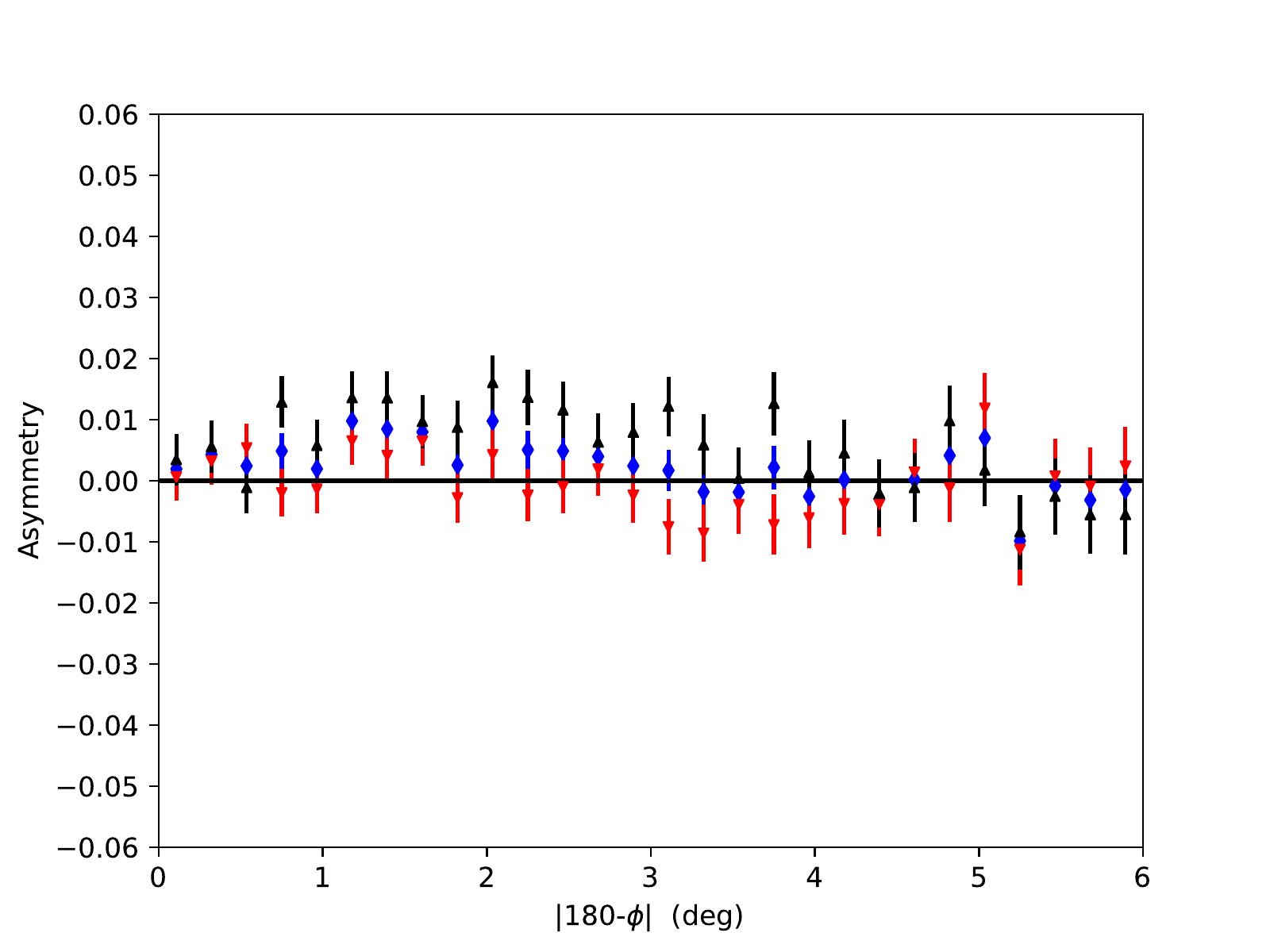}}
\subfloat[]{\includegraphics[scale=0.55]{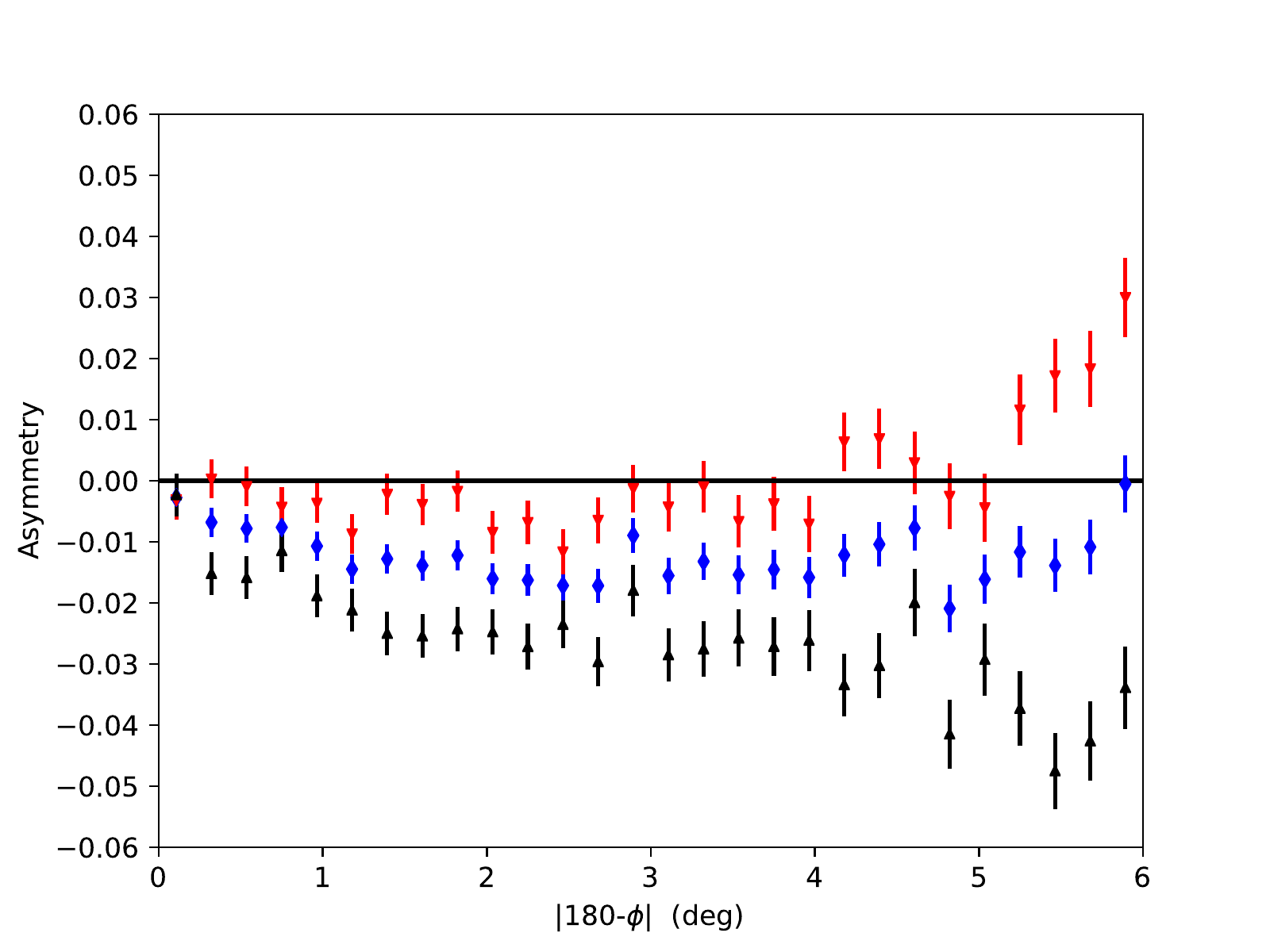}}

\subfloat[]{\includegraphics[scale=0.55]{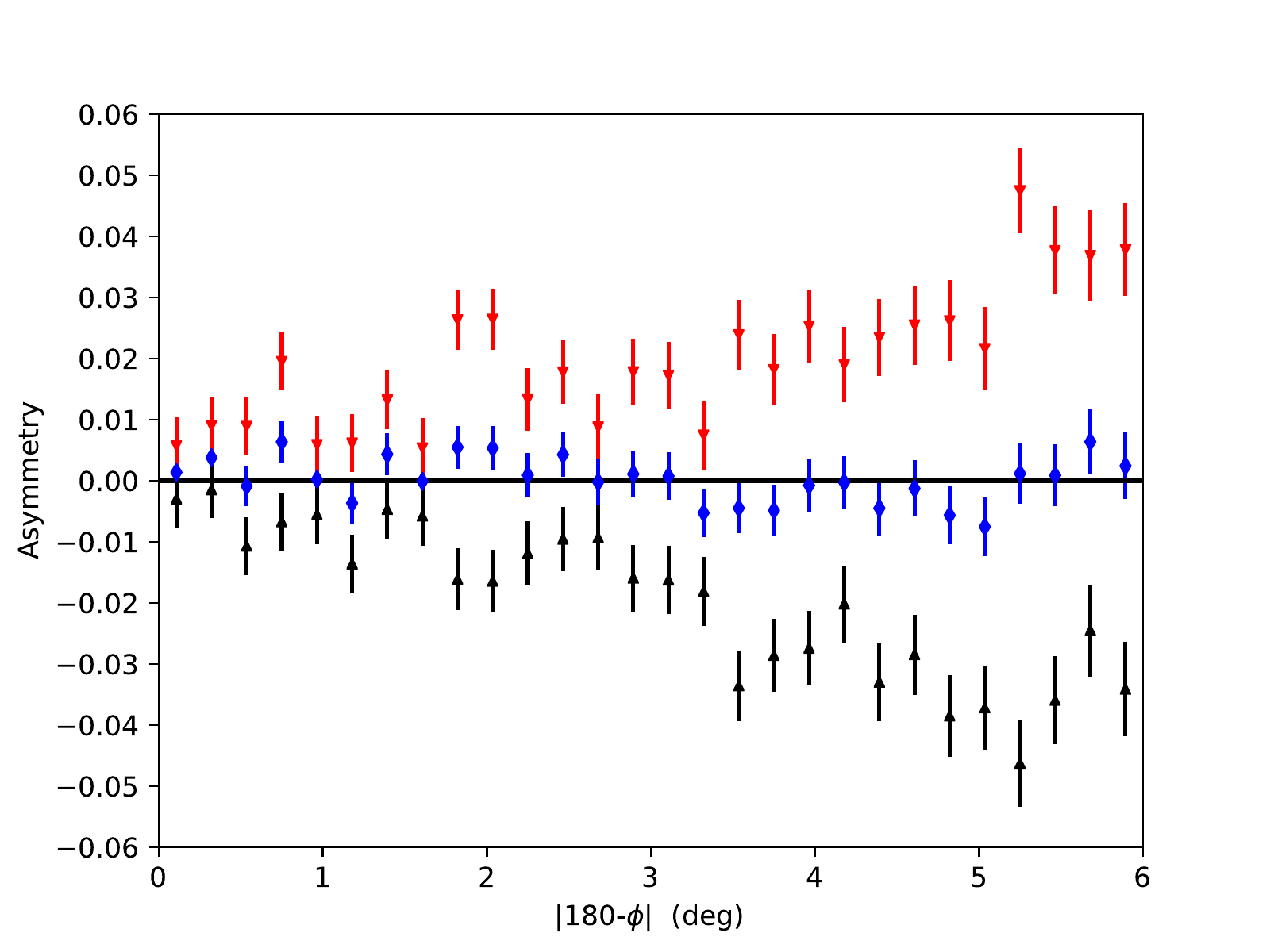}}
\caption{
The radial study of the data set which include a stricter faintness limit of $G < 17$.
(a)  The axial symmetry test for stars with $R \in [7,7.7]$ kpc.  
(b)  The axial symmetry test for stars with $R \in [7.7,8.3]$ kpc.  
(c)  The axial symmetry test for stars with $R \in [8.3,9]$ kpc.  
Additional cuts used in all panels are: $0.5 < G_{\rm BP} - G_{\rm RP} < 2.5$ mag, $14 < G < 17$ mag, $\varpi > 0$ mas, $|b| > 30^{\circ}$, $0.2 < |z| < 3.0$ kpc, $|180^{\circ} - \phi| \leq 6^{\circ}$, and the LMC/SMC excision outlined in Eqs.~\ref{LMCcuts} and \ref{SMCcuts}.
}
\label{fig:Rdepend_Gcut}
\end{center}
\end{figure}

For the first 
method, we see from Fig.~\ref{fig:Rdepend_Gcut} that the two outermost radial bins do not change appreciably, which matches what one would expect given that the crowded regions of the sky do not subtend a large portion of the geometry there.  For the innermost $R$ bin in panel a), however, the GC
is 
behind nearly all of the data.  Minding the ``hazy" G-band magnitude completeness limits of  the {\it Gaia} telescope \citep{lindegren2018gaia}, we cut out the dimmest stars in order to test whether or not incompleteness affects the data.  The N+S (blue diamonds) data does not change appreciably, while the N and S only curves become only slightly closer to zero.

\begin{figure}[ht!]
\begin{center}
\subfloat[]{\includegraphics[scale=0.55]{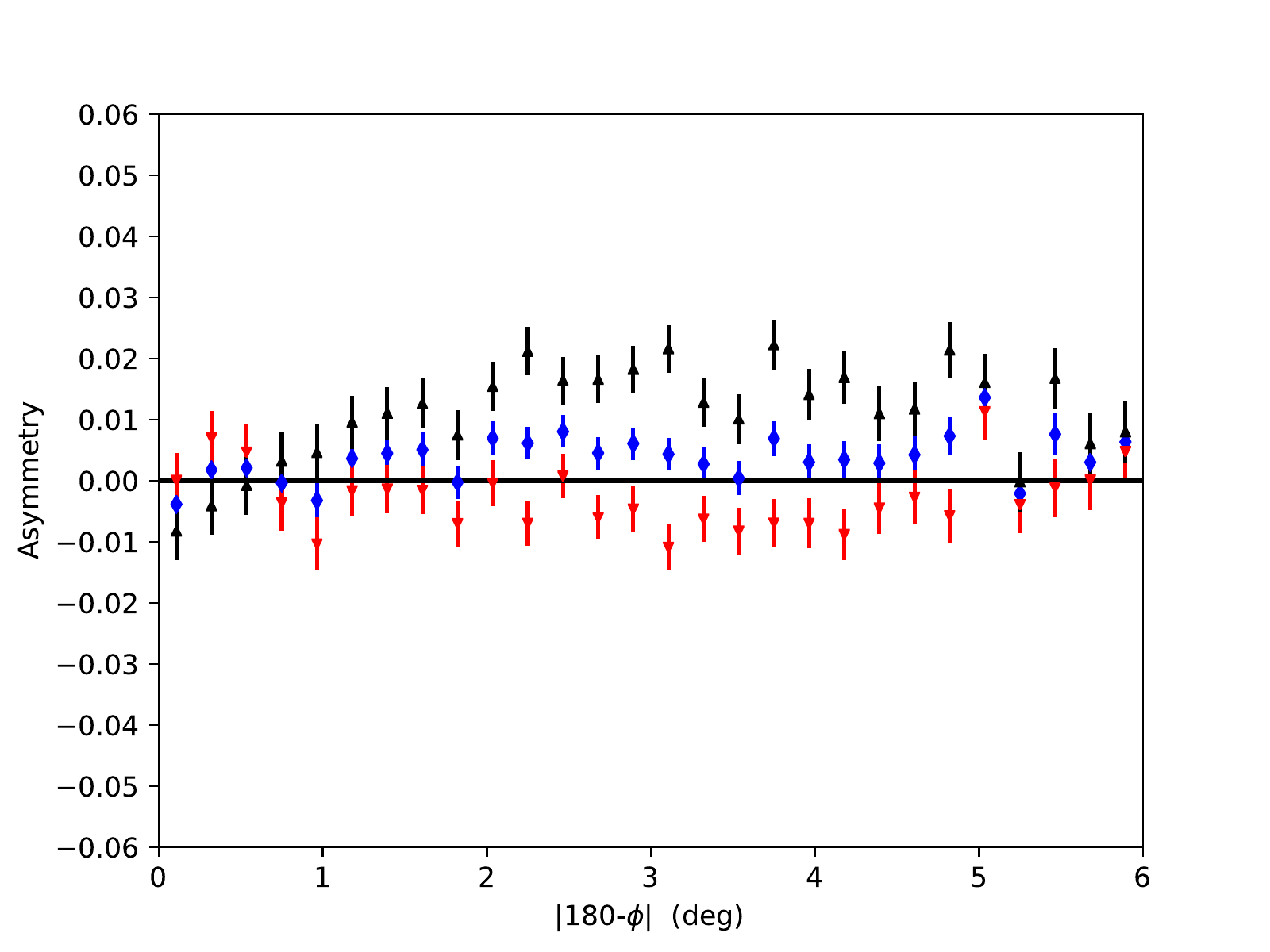}}
\subfloat[]{\includegraphics[scale=0.55]{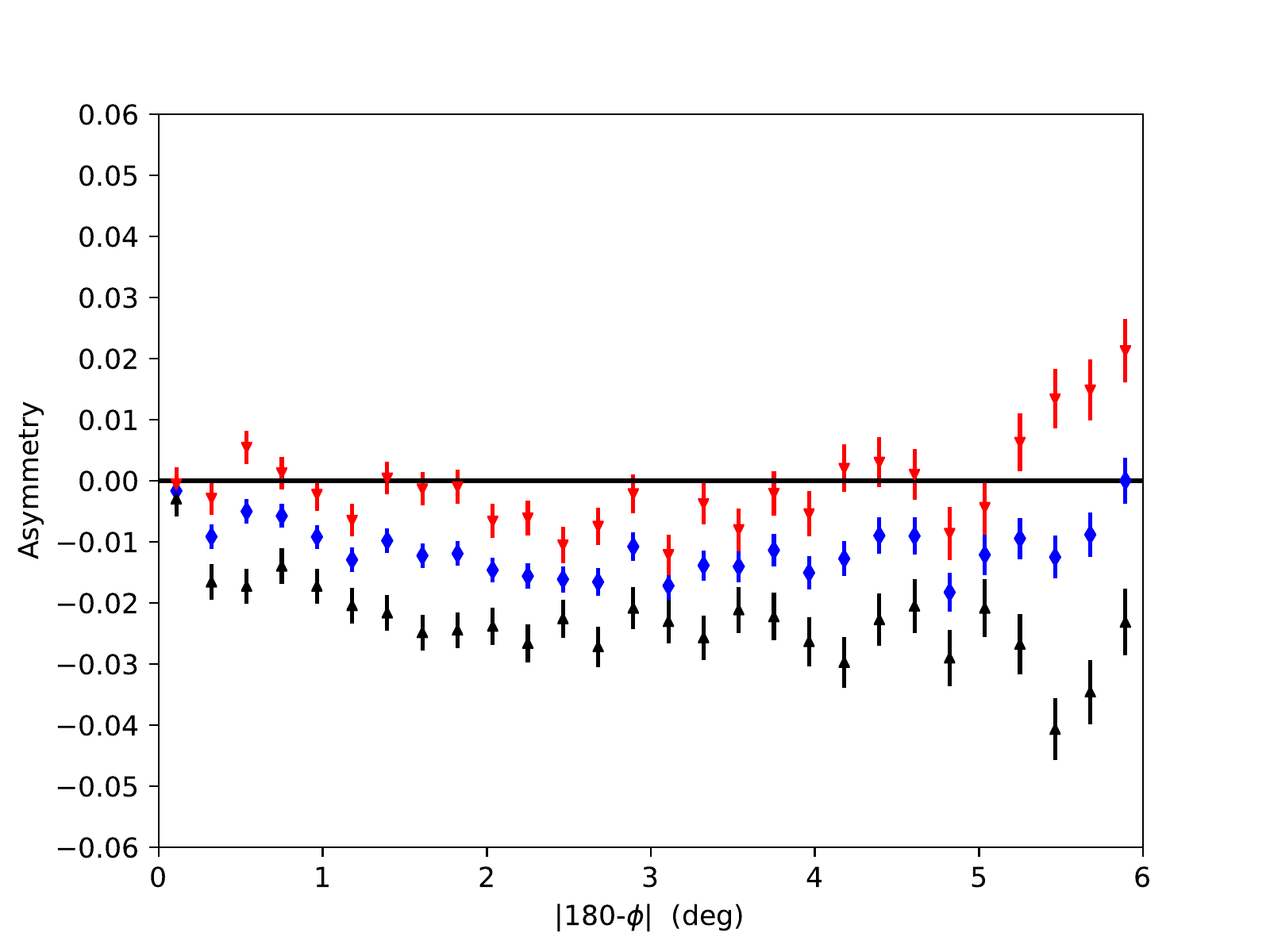}}

\subfloat[]{\includegraphics[scale=0.55]{Rstudy/FINAL_azasym_G1418_R8390_color0525_z0230_prlxnewLMCSMC.pdf}}
\subfloat[]{\includegraphics[scale=0.45]{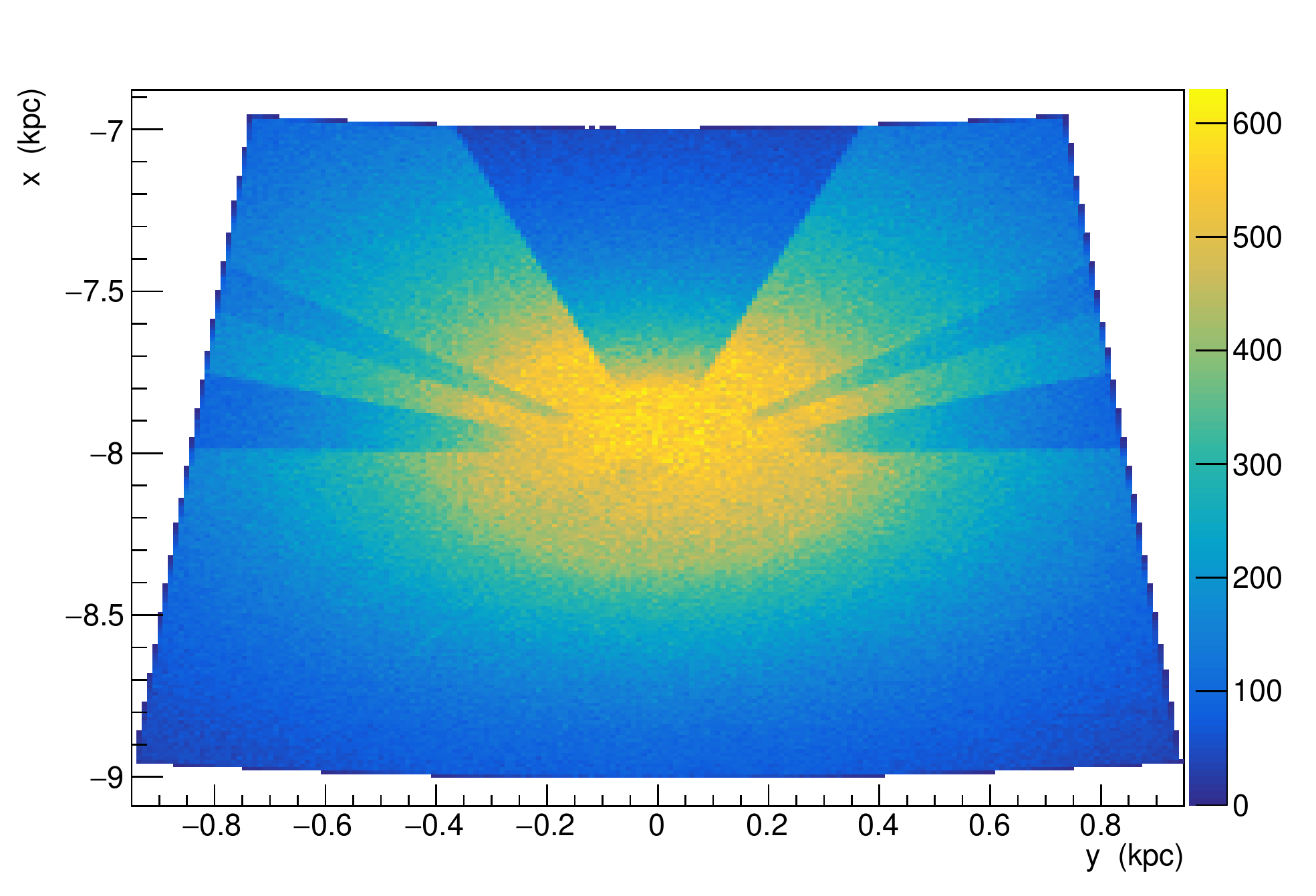}}
\caption{
The radial study of the data set which has had the 
GC excised.
(a)  The axial symmetry test for stars with $R \in [7,7.7]$ kpc.  
(b)  The axial symmetry test for stars with $R \in [7.7,8.3]$ kpc.  
(c)  The axial symmetry test for stars with $R \in [8.3,9]$ kpc is not affected by the GC excision and thus this is the same plot as Fig.~\ref{fig:Rdepend_normal} c).    
(d)  The $xy$ footprint of the data when the GC excision is included.  
Additional cuts used in all panels are: $0.5 < G_{\rm BP} - G_{\rm RP} < 2.5$ mag, $14 < G < 18$ mag, $\varpi > 0$ mas, $|b| > 30^{\circ}$, $0.2 < |z| < 3.0$ kpc, $|180^{\circ} - \phi| \leq 6^{\circ}$, and the LMC/SMC excision outlined in Eqs.~\ref{LMCcuts} and \ref{SMCcuts}.
}
\label{fig:Rdepend_lbcuts}
\end{center}
\end{figure}

For the second method, we choose to excise the densest region of the $l,b$ plot in Fig.~\ref{fig:lbcuts} b).  To wit, we cut out all stars within $20^{\circ}$ of the $l=0^{\circ}$ line, which also satisfy $|b| < 45^{\circ}$.  The results of this check are shown in Fig.~\ref{fig:Rdepend_lbcuts}, where it is apparent that there is no appreciable change in any of the bins.  Additionally, Fig.~\ref{fig:Rdepend_lbcuts} d) shows the altered $xy$ footprint with the GC-masking $l,b$ cuts employed.

\begin{figure}[ht!]
\begin{center}
\subfloat[]{\includegraphics[scale=0.55]{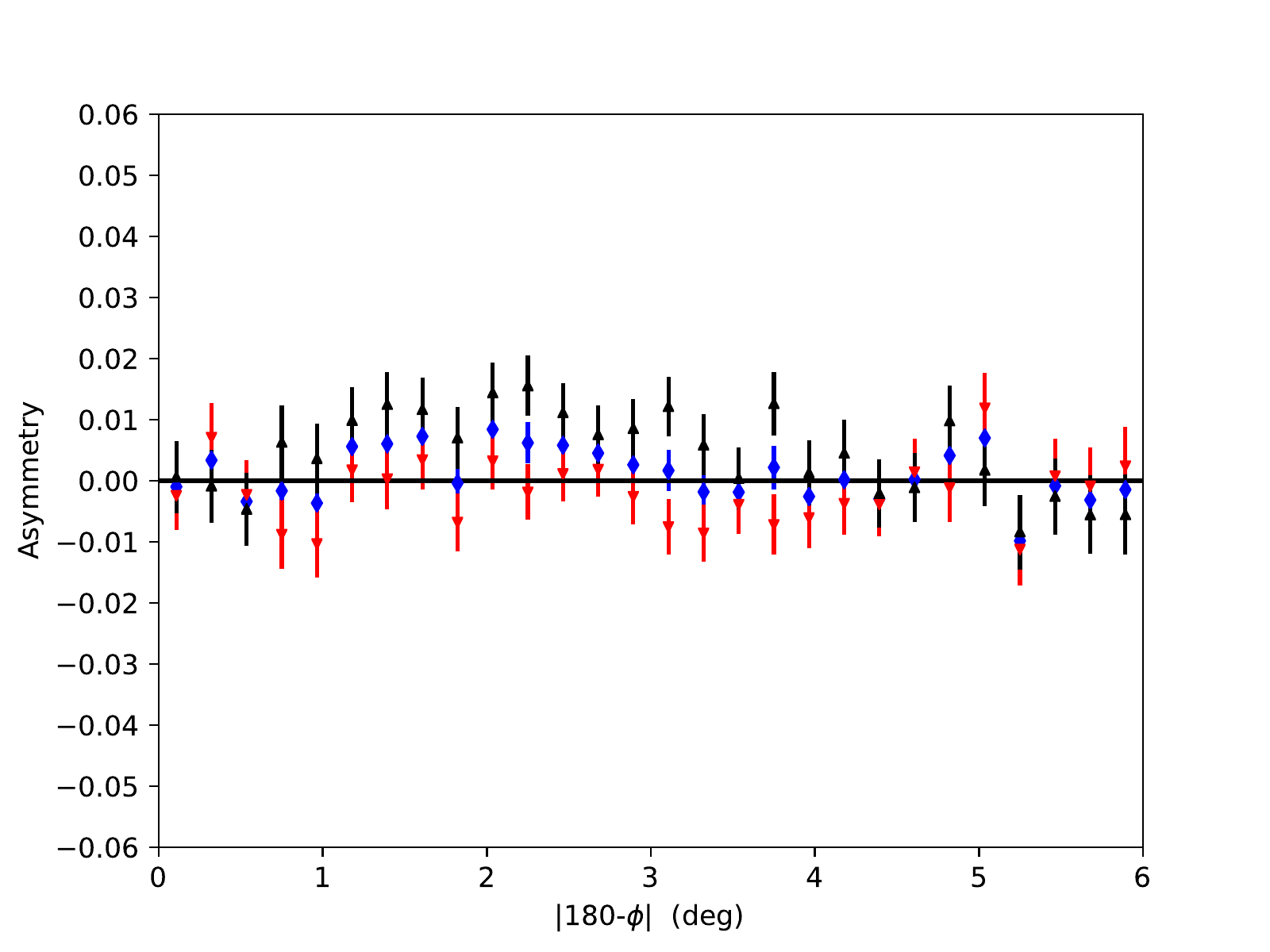}}
\subfloat[]{\includegraphics[scale=0.55]{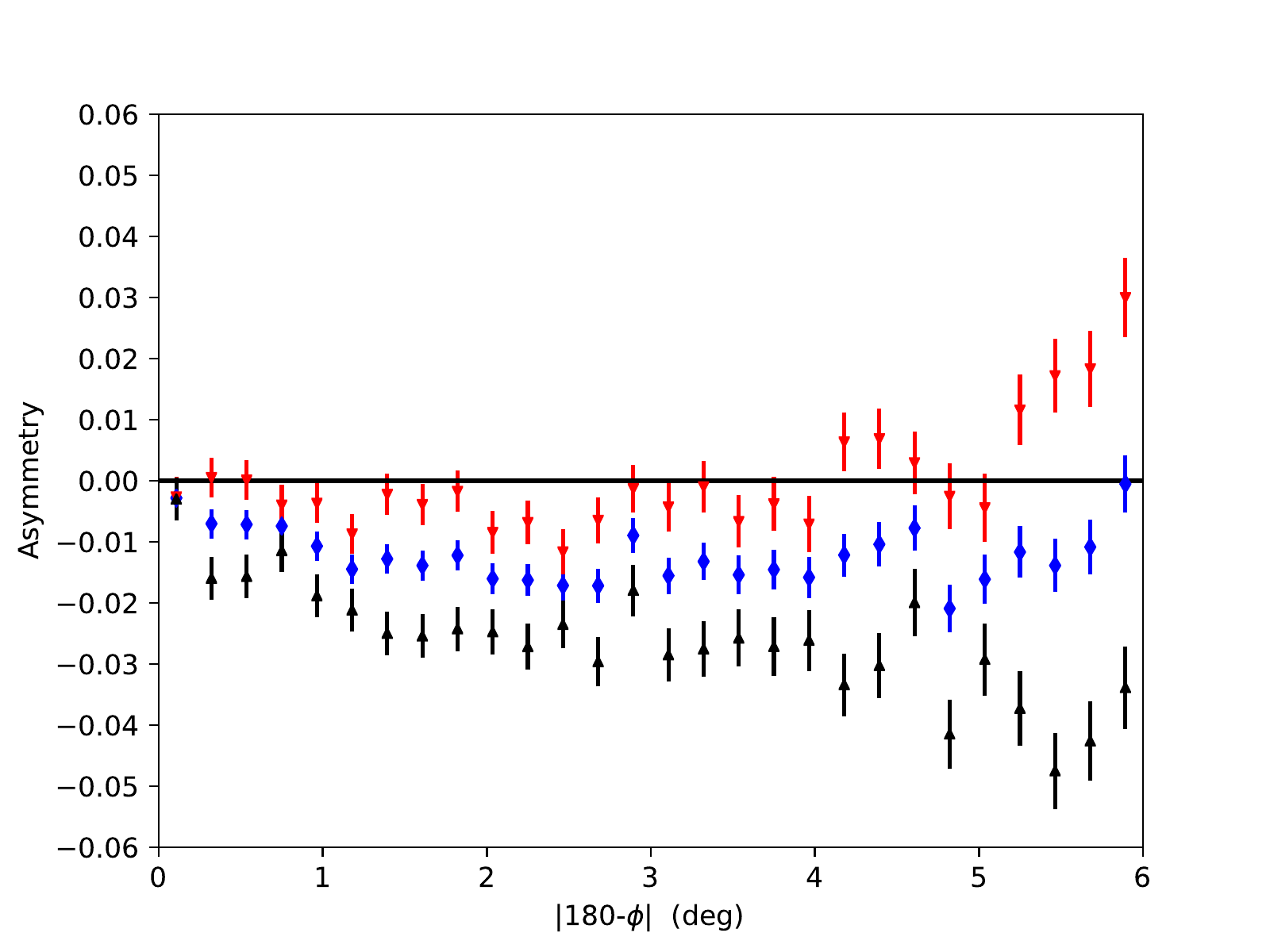}}

\subfloat[]{\includegraphics[scale=0.55]{Rstudy/FINAL_azasym_G1417_R830900_color0525_z0230_prlxnewLMCSMC.pdf}}
\caption{
The radial study of the data set which include a stricter faintness limit of $G < 17$ mag and the GC  
excision.
(a)  The axial symmetry test for stars with $R \in [7,7.7]$ kpc.  
(b)  The axial symmetry test for stars with $R \in [7.7,8.3]$ kpc.  
(c)  The axial symmetry test for stars with $R \in [8.3,9]$ kpc is not affected by the GC excision and thus this is the same plot as Fig.~\ref{fig:Rdepend_Gcut} c).  
Additional cuts used in all panels are: $0.5 < G_{\rm BP} - G_{\rm RP} < 2.5$ mag, $14 < G < 17$ mag, $\varpi > 0$ mas, $|b| > 30^{\circ}$, $0.2 < |z| < 3.0$ kpc, $|180^{\circ} - \phi| \leq 6^{\circ}$, and the LMC/SMC excision outlined in Eqs.~\ref{LMCcuts} and \ref{SMCcuts}.
}
\label{fig:Rdepend_Gandlbcuts}
\end{center}
\end{figure}

For the third and final method of checking that our result is not appreciably afflicted by completeness and stellar identification issues in the dense GC 
region, we implement both the magnitude and ``box" cuts mentioned above and show these results in Fig.~\ref{fig:Rdepend_Gandlbcuts}.  Clearly, the largest $R$ bin is unaffected by the $l,b$ cuts and is identical to Fig.~\ref{fig:Rdepend_Gcut} c).  The intermediate $R$ bin (Fig.~\ref{fig:Rdepend_Gandlbcuts} b)) shows no appreciable change.  Interestingly, the innermost $R$ bin appears to exhibit a wave-like N+S asymmetry\footnote{ Although the 
asymmetries associated with innermost and outermost radial 
bins of Fig.~\ref{fig:Rdepend_Gandlbcuts} appear to suggest 
wave-like features, a reduced chi-squared analysis reveals that Fig. 11 a) contains the only 
statistically significant
wave-like effect. }, a feature that was not as easily identifiable with the likely 
more poorly measured 
GC-direction stars and dim stars included.  Otherwise, the general trend remains mostly similar to the main result and the other tests.

\subsection{Comparison of stricter cuts on the aggregate asymmetry result}

\begin{figure}[ht!]
\begin{center}
\subfloat[]{\includegraphics[scale=0.55]{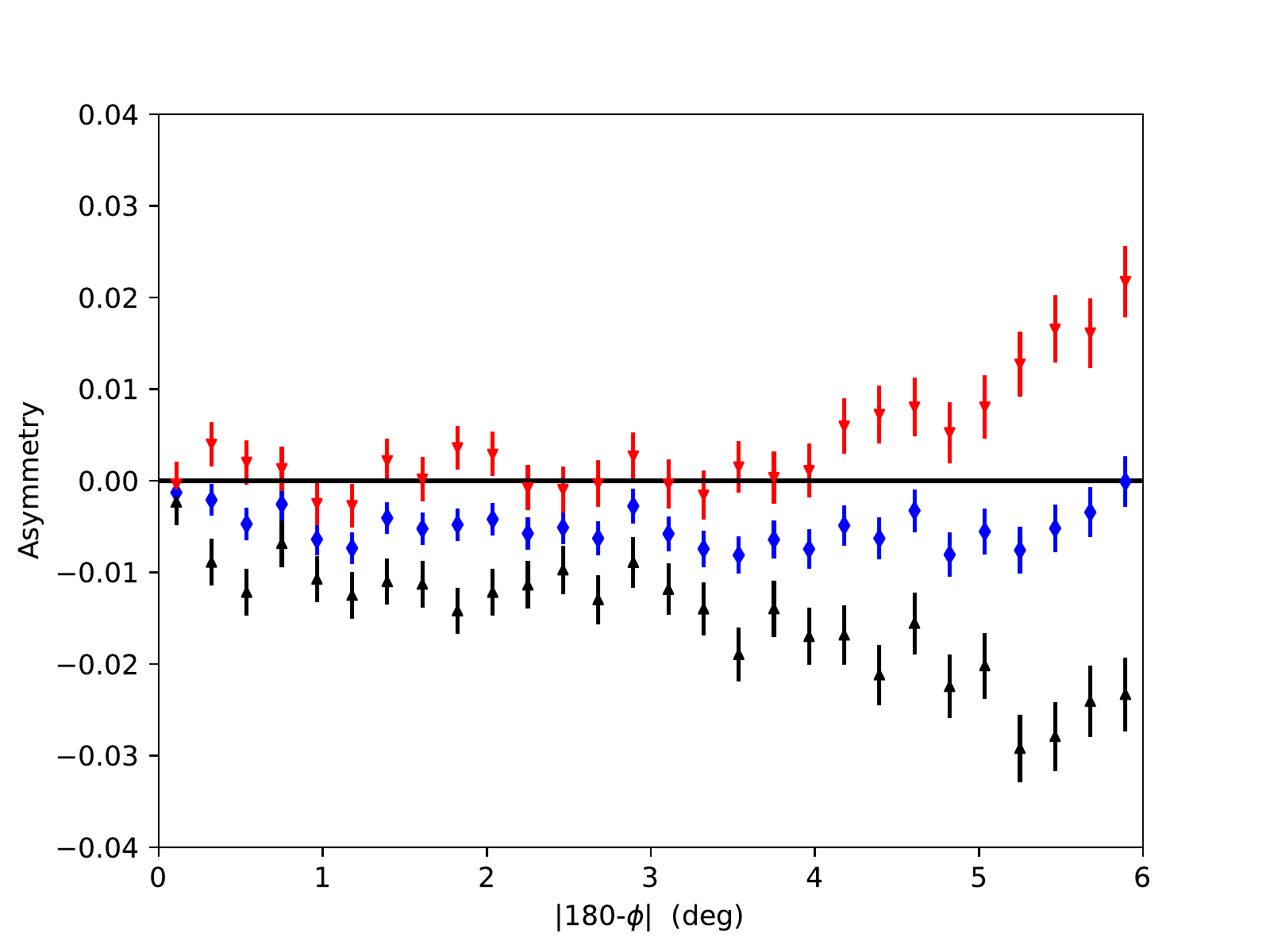}} 
\subfloat[]{\includegraphics[scale=0.55]{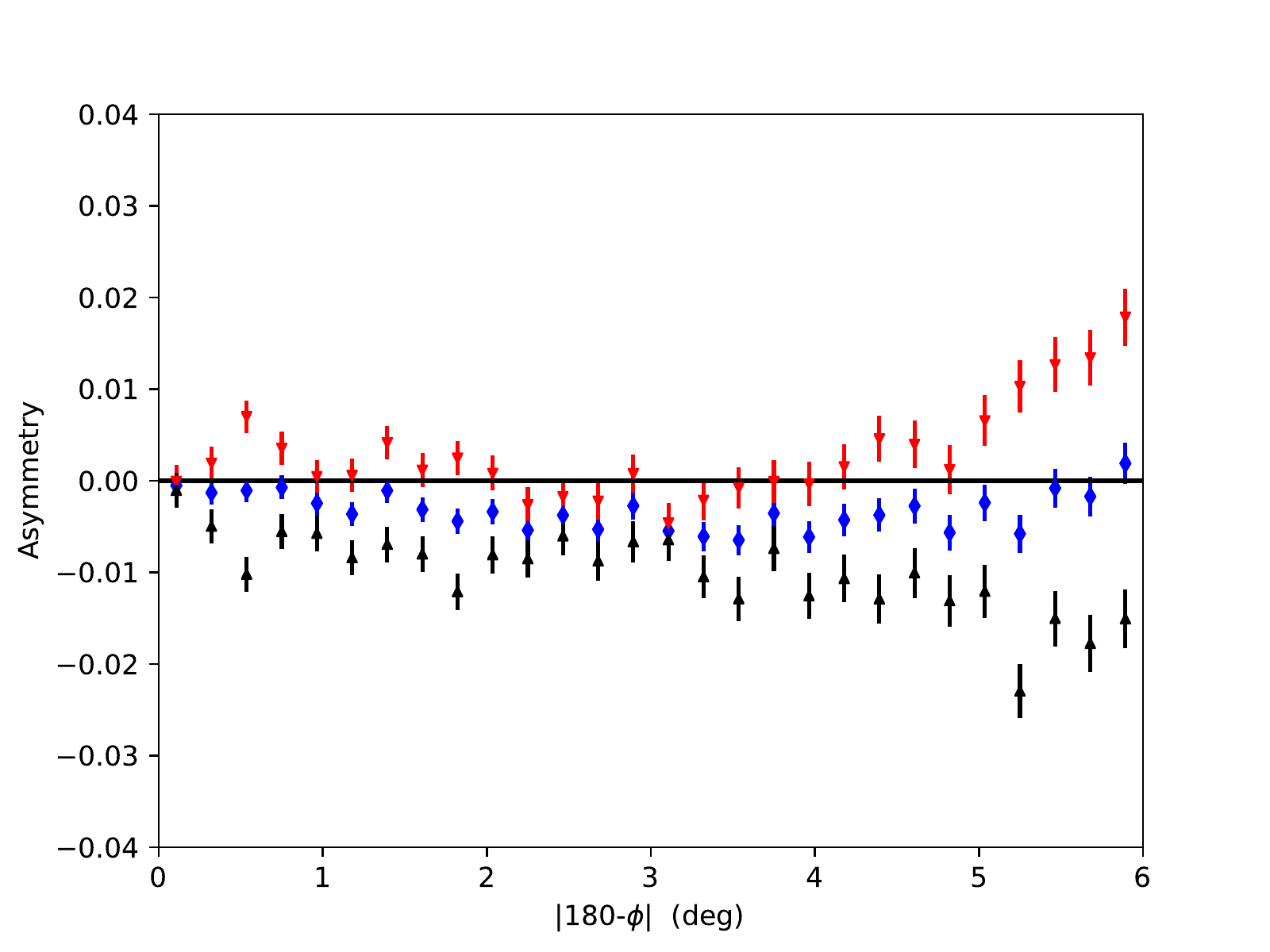}}

\subfloat[]{\includegraphics[scale=0.55]{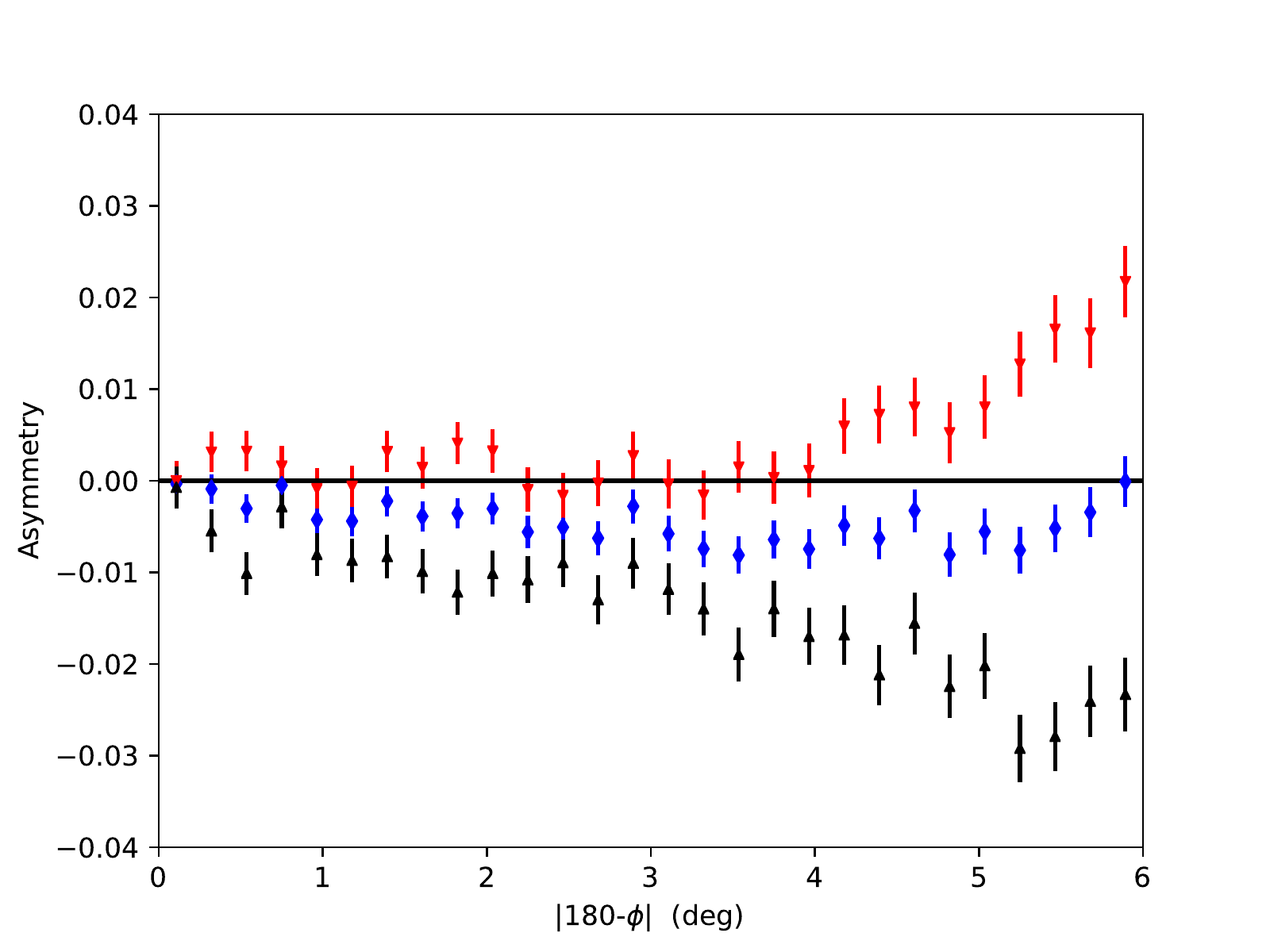}}
\caption{
(a)  Aggregate study of axial symmetry with $G < 17$ mag and the 
GC cuts of the above tests.  
(b)  Aggregate study of axial symmetry from \citet{GHY20} with $G < 18$ mag.
(c)  The same as panel (b), but with $G<17$ mag.
Additional cuts used in all panels are: $0.5 < G_{\rm BP} - G_{\rm RP} < 2.5$ mag, $\varpi > 0$ mas, $|b| > 30^{\circ}$, $0.2 < |z| < 3.0$ kpc, $|180^{\circ} - \phi| \leq 6^{\circ}$, and the LMC/SMC excision outlined in Eqs.~\ref{LMCcuts} and \ref{SMCcuts}.
}
\label{fig:Aggregate}
\end{center}
\end{figure}

In order to assess if the additional cuts outlined in \S~\ref{subsec:GCcuts} appreciably change the resulting, aggregate asymmetry found in \citet{GHY20}, we apply the same 
GC masking to the aggregate data set in Fig.~\ref{fig:Aggregate}.  Clearly, the results do not change qualitatively at all, and there is only a negligible quantitative change in the signal we find for each set of cuts.  This fact allows us to conclude that potentially poorly measured stars in the denser regions of the sky near the 
GC are not responsible for the asymmetry we see.

\subsection{Red and blue color cut analysis}

\begin{figure}[ht!]
\begin{center}
\subfloat[]{\includegraphics[scale=0.55]{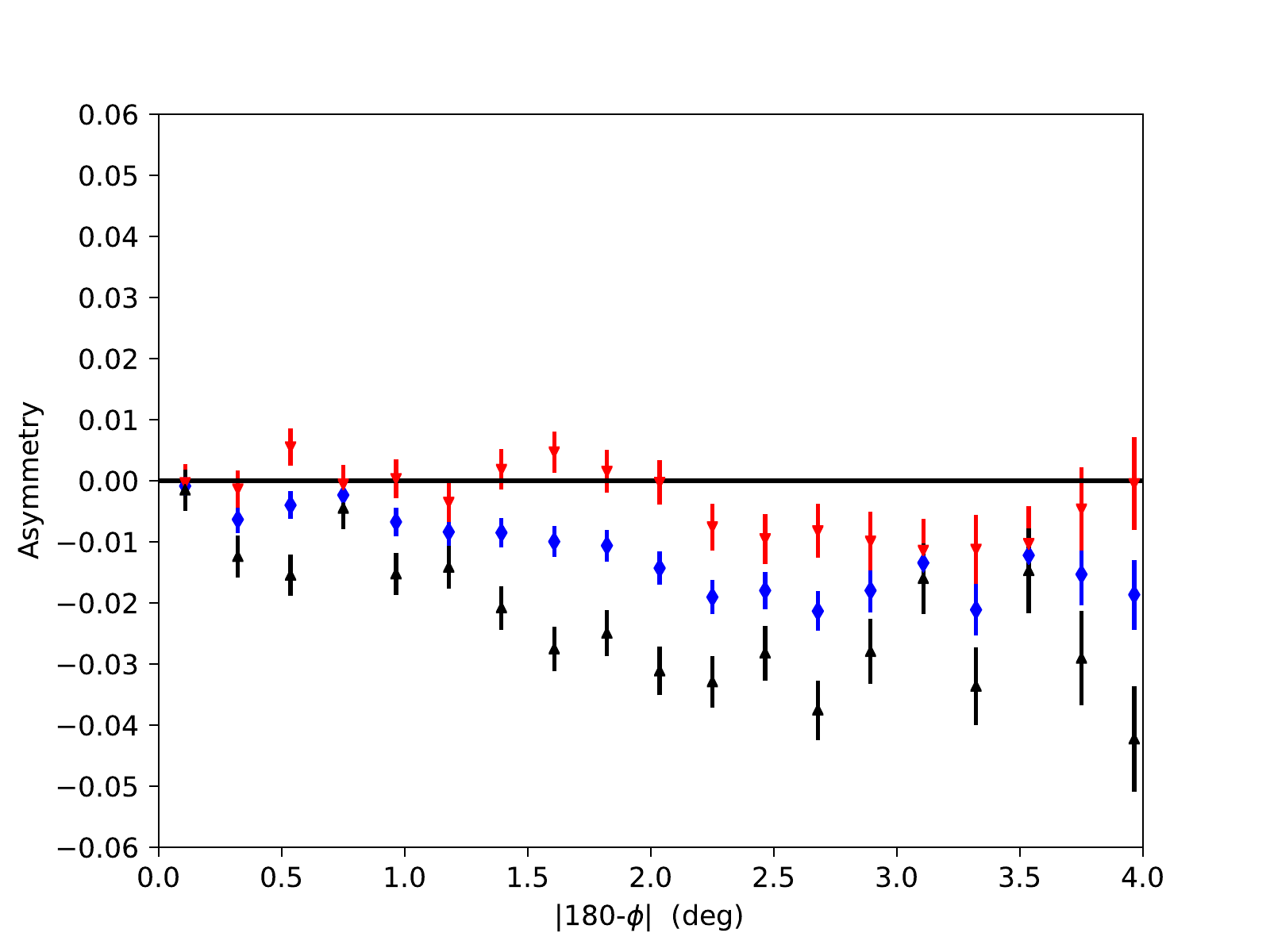}}
\subfloat[]{\includegraphics[scale=0.55]{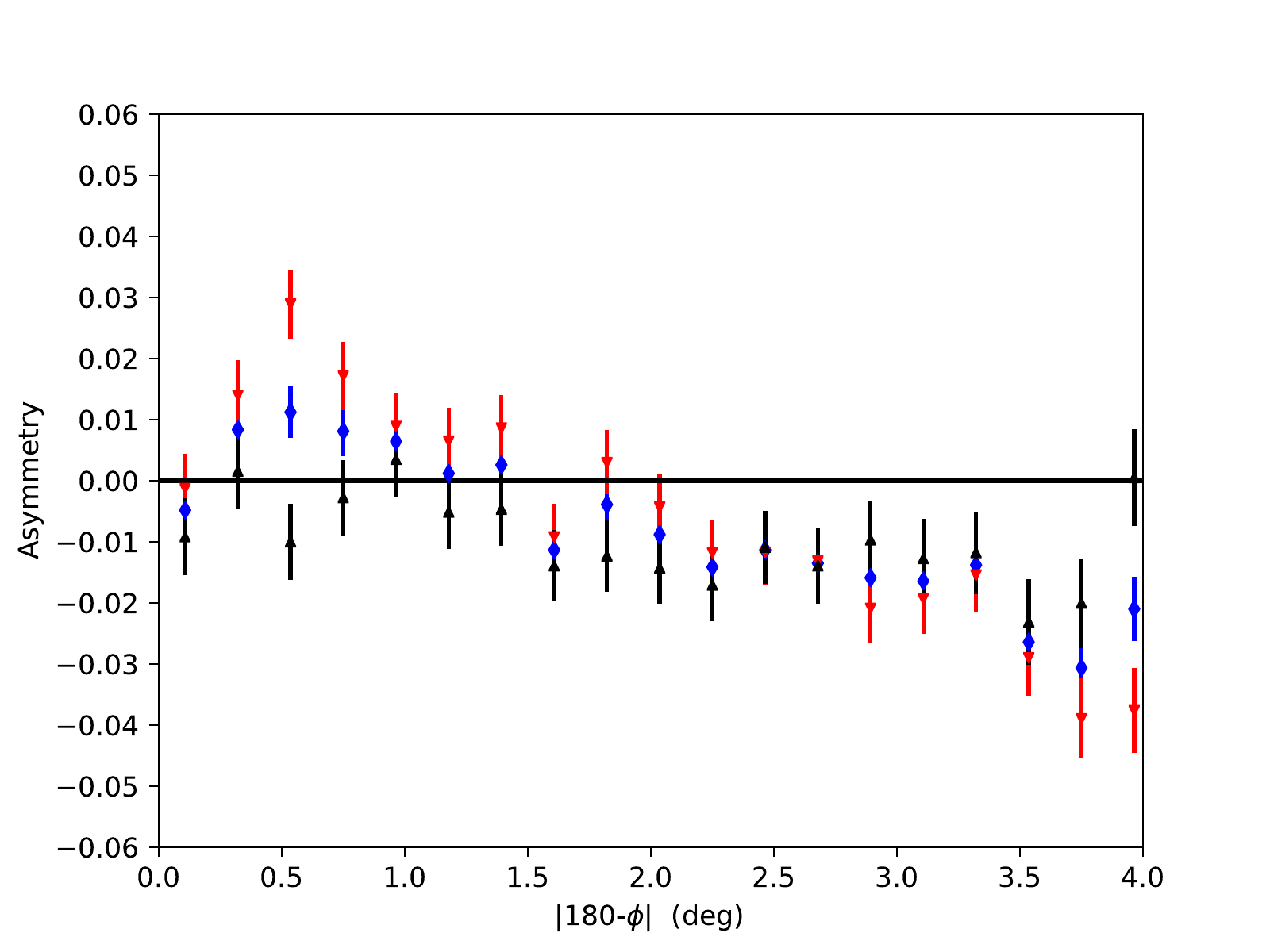}}
\caption{
(a) Test of axial symmetry with red ($1.5 < G_{\rm BP} - G_{\rm RP} < 2.5$ mag) stars only over a small volume of nearby space: $0.2 < |z| < 0.5$ kpc and $R \in [7.5,8.5]$ kpc. 
(b) Test of axial symmetry with blue ($0.5 < G_{\rm BP} - G_{\rm RP} < 1.5$ mag) stars only over the same volume of space as panel a).  
Additional cuts used in both panels are:  $\varpi > 0$ mas, $|b| > 30^{\circ}$, $|180^{\circ} - \phi| \leq 4^{\circ}$, and the LMC/SMC excision outlined in Eqs.~\ref{LMCcuts} and \ref{SMCcuts}.
}
\label{fig:axialwithcolor}
\end{center}
\end{figure}

Using only the nearest volume subset of our sample in order to isolate color-dependence (see Fig.~\ref{fig:axialwithcolor} caption for cuts), we see that the behavior of the red stars (Fig.~\ref{fig:axialwithcolor}a) and the blue stars (Fig.~\ref{fig:axialwithcolor}b) are both consistent with a downward trend in the N+S (blue diamonds) data, lending support to an overall distortion of the shape of the potential which is affecting all stars.  There are, however, slight differences between the red and blue stars for the N-S data.  This may be due to some effect seen only in the older, redder population, but we do not speculate on the specific cause.

\subsection{Height above the plane analysis}

\begin{figure}[ht!]
\begin{center}
\subfloat[]{\includegraphics[scale=0.55]{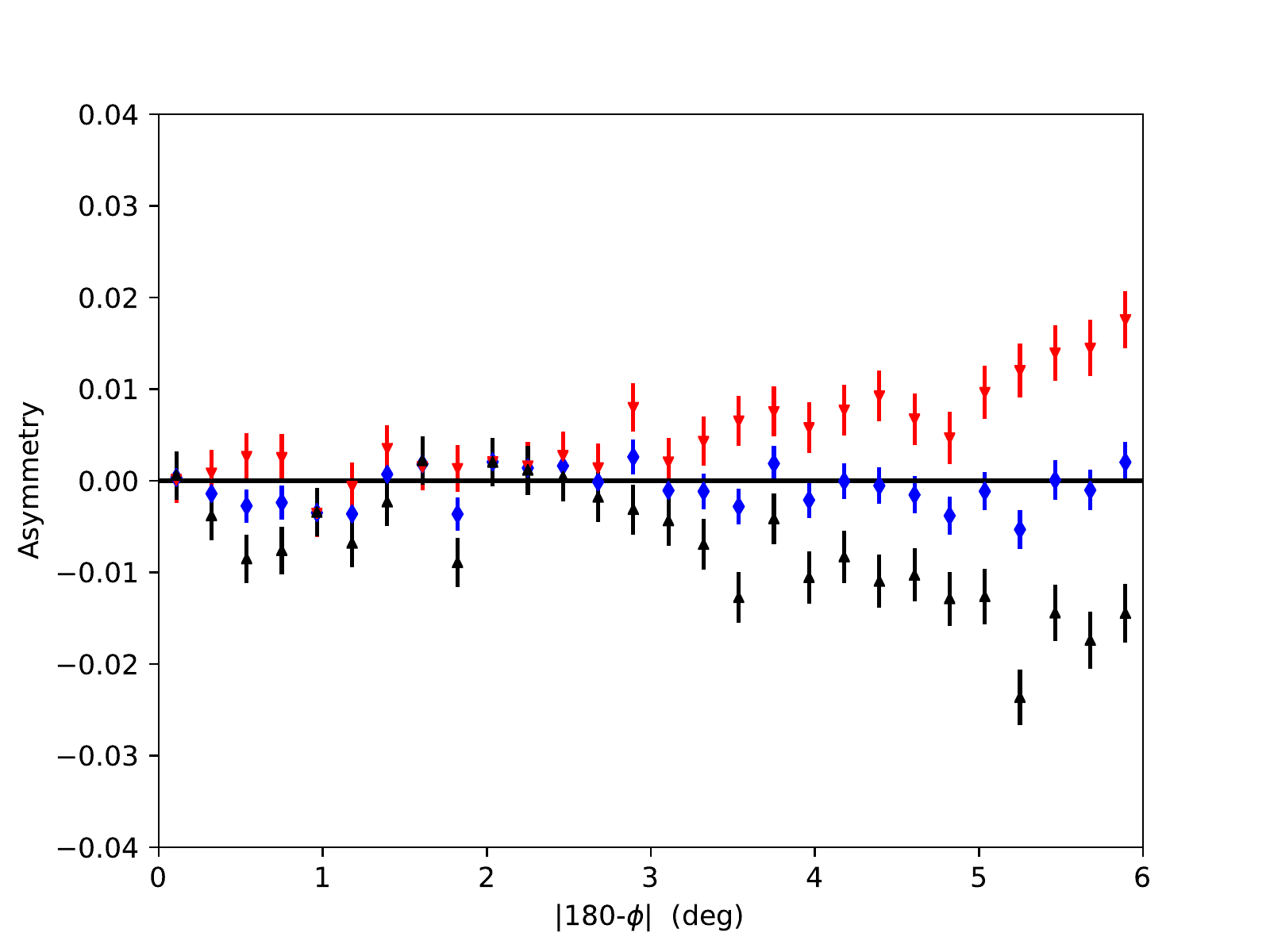}}
\subfloat[]{\includegraphics[scale=0.55]{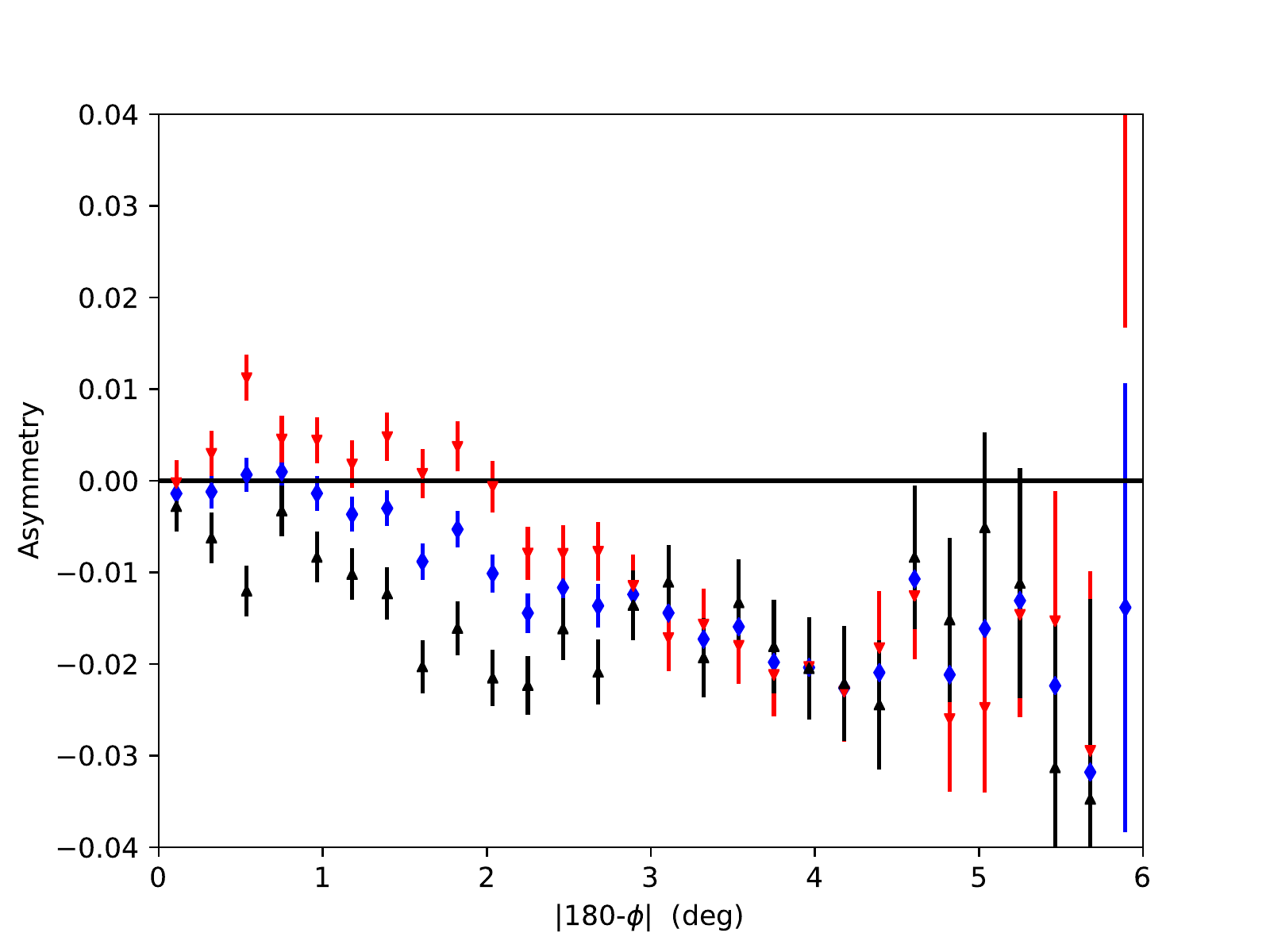}}
\caption{
(a)  Far from plane sample, $0.5 < |z| < 3$ kpc.  
(b)  Close to plane sample, $0.2 < |z| < 0.5$ kpc. 
Additional cuts used in both panels are: $0.5 < G_{\rm BP} - G_{\rm RP} < 2.5$ mag, $\varpi > 0$ mas, $|b| > 30^{\circ}$, $7 < R < 9$ kpc, $|180^{\circ} - \phi| \leq 6^{\circ}$, and the LMC/SMC excision outlined in Eqs.~\ref{LMCcuts} and \ref{SMCcuts}. As the $|b|$ cuts cause these panels to sample very different swaths of $R$, 
this figure does not show $z$-dependence alone. The low-$z$ stars preferentially sample the region of $R$ 
close to 
the Sun due to the geometry of the latitude cuts, which is consistent with the asymmetry expected from the region just beyond the OLR.}

\label{fig:Zdepend}
\end{center}
\end{figure}

Finally, upon dividing the sample into subsets above and below $|z| = 0.5$ kpc, as in Figs.~\ref{fig:Zdepend}
a) and b), respectively, one notices that the downward trend in the N+S data set (blue diamonds) appears to be much more marked for the low $|z|$ stars, seemingly revealing 
behavior at odds with 
what one would expect if a distorted halo were the main cause of the asymmetry. 
However, we have carefully removed the spiral arms 
via our $b$ and $z$ cuts.  
We note, however, that the low-$z$ stars preferentially sample the region of $R$ close to the Sun due to the geometry of the latitude cuts imposed on the sample.
Indeed, this behavior is completely consistent with a picture 
in which the OLR is inside the solar circle, as one would expect the low-$|z|$ stars near $R = R_{0}$ to exhibit negative values of the asymmetry, due to the tendency of orbits to align with the bar just outside of the OLR \citep{binney2008GD}. In contrast, the more marked N-S difference for the high-$|z|$ stars is consistent with a tilted prolate halo.

\section{Results}

As evident from Figs.~\ref{fig:Aggregate} and 
\ref{fig:Rdepend_normal}, our results do 
confirm the findings of \citet{GHY20}, interpreted as an overall effect due to the LMC \& SMC system, 
but also reveal $R$-dependent features.  
In particular, 
the sign flip in the asymmetry in Fig.~\ref{fig:Rdepend_normal} as one looks closer to the GC 
suggests that another object contributes in that region.  Given that the second most significant perturber noted by \citet{GHY20} is the Galactic bar, and that the signal occurs at smaller  $R$, 
the Galactic bar is the most likely 
culprit.  

In fact, due to the non-axisymmetric, time-dependent nature of the Galactic bar potential, resonances can occur at very specific galactocentric radii \citep{binney2008GD}.  While the existence of a family of stars in bar-resonant orbits depends strongly on the relative strength of the Galactic bar, it is the case for all but the strongest of bars that resonant orbits between the Co-rotation Resonance (CR) and the Outer Lindblad Resonance (OLR) orbit with trajectories perpendicular to the bar, and beyond the OLR the resonant orbits tend to be elongated along the bar's orientation \citep{contopoulos1980orbits, binney2008GD}.  This non-axisymmetric behavior could easily be responsible for the sign-flip we see in Fig.~\ref{fig:Rdepend_normal}, as there has long been a notion that a resonance occurs just interior to the solar circle \citep[e.g.]{mishurov1999yes, dehnen2000effect}. Interestingly, 
given that the location of the Galactic bar is known, we believe
that the pattern of axial asymmetries we see, that ${\cal A}$ flips
sign from positive to negative, rather than from negative to positive, 
indicates that the effect is an Outer Lindblad Resonance. 

Finally, given the importance of Sagittarius in exciting vertical motions in the Galactic disk \citep{widrow2012galactoseismology,yanny2013stellar,laporte2018influence}, one might ask if it, too, could have an effect axially on our sample.  However, due to the alignment of the perturber and stars in our sample, the $z$-component of the torque would be very small, as the displacement vector and force vector are very nearly aligned, yielding a severely less significant torque. We expand upon the possible impact
of the evolutionary history of the Sgr dwarf and stream below. 

Thus the axial asymmetry profiles we have found appear 
consistent 
with two overarching effects: 
first, a matter distribution warped by the LMC \& SMC, 
as suggested by the 
distorted halo model determined from studies of the Orphan stream by \citet{erkal2019total}, and second, an asymmetry associated with an 
Outer Lindblad Resonance driven by the Galactic bar. 

In the distorted halo model, the dark halo of the Milky Way is stretched by the LMC/SMC system, causing an elongation in the direction of the Magellanic clouds \citep{erkal2019total}.  With the Miyamoto-Nagai disk \citep{miyamoto1975disk} and distorted NFW halo profile \citep{navarro1997universal} of \citet{erkal2019total} integrated over the same region as our data set, the N only, S only, and N+S data sets 
all qualitatively match a reflex prolate distribution much better than the other halo geometries offered \citep{GHY20}. 
As we have mentioned in 
\citet{GHY20}, this halo geometry is also consistent with the galactic warp picture of \citet{weinberg2006magellanic} and \citet{dekel1983galactic}.

It is perhaps disconcerting, though, that upon dividing the data set into radial or vertical bins, the allegedly global asymmetry effect of a distorted 
matter distribution 
is not seen at the same level in each bin.  Indeed, the vertical-axial correlation and N+S trend of Fig.~\ref{fig:Rdepend_normal} a) is reversed from what is expected, while panels b) and c) still fit well with this picture.  Additionally, as can be seen from Fig.~\ref{fig:Zdepend} the axial asymmetry is largely seen in stars close to the plane, the opposite of what one would expect for a halo-driven 
effect.  However, due to the nature of the $|b|$ cuts implemented here, different bands of $|z|$ sample different values of $R$, as can be understood from Fig.~\ref{fig:geometry} b).  Thus, it is possible that the $z$-dependence is confounded by sampling different values of $R$.  Regardless, the picture of Magellanic cloud influence alone cannot explain the innermost $R$ bin, and thus requires some additional effect.  This effect is likely from the Galactic Bar as it is only seen at the innermost $R$ bin. 
Indeed we believe 
it to be the result of the presence of the Outer Lindblad Resonance (OLR). This effect, believed to be just inside the solar circle \citep{dehnen2000effect}, can cause perturbations to stars just inside (outside) the OLR which causes them to align themselves in a perpendicular (parallel) sense with respect to the bar \citep{contopoulos1980orbits, dehnen2000effect}.  Given that the Galactic Bar is known to point at approximately $13 - 27^{\circ}$ from the Sun's location \citep{robin2012stellar, portail2016structure}, this orbital alignment configuration would give rise to an axial asymmetry which is Left-heavy (${\cal A} > 0$) for values of $R$ just inside the OLR, and Right-heavy (${\cal A} < 0$) for values of $R$ just outside the OLR, and no net effect well beyond the OLR.  This qualitatively matches the results of Fig.~\ref{fig:Rdepend_normal}. Note that if the effect
were a Co-Rotation resonance, the asymmetry ought to flip from negative to positive. 

It is unclear exactly how the stars near the OLR can behave in such a way to also cause a N-S asymmetry, but it has been noted that the vertical resonance from a central bar can have significant effects on the vertical amplitudes of orbits near the Inner Lindblad Resonance \citep{quillen2002growth, hasan1993galactic}, so that such an effect near the OLR could well also occur. 

Upon consideration of the above effects, it seems most likely that both Magellanic torque and bar induced resonances are operative in the local star count data.  The OLR creates a Left-heavy asymmetry in the innermost $R$ bin and constructively interferes with the 
LMC \& SMC-driven (halo) effect in the middle $R$ bin.  Finally, the OLR ceases to play a large role at higher $R$ and the 
halo 
effect is the lone cause of the signal we see.

We have determined the location of the in-plane radius at which the axial asymmetry, N+S, 
flips sign as follows. We compute the 
asymmetry 
$\langle {\cal A (\phi) } \rangle$, making an error-weighted average over azimuthal angles such that 
$|180^{\circ}-\phi| < 6^{\circ}$, where ${\cal A}(\phi)$ itself 
counts up stars in 
a wedge of width $\Delta R$, for various
choices of starting radius $R_i$. We refine the determined radius at which $\langle {\cal A}(\phi) \rangle$
flips sign through an iterative procedure.
That is, we begin with wedges of 
width $\Delta R = 500$ pc and move outward in $R_i$ by 200 pc increments, to determine 
where the average asymmetry 
changes sign.  After observing a sign flip, the radial width of each bin is decreased, and the scan repeated over a smaller range of $R_i$ 
to sharpen the determination of the radial location of the sign flip. 
The uncertainty is derived from the smallest radial wedge, of 
200 pc,  
which still reveals a sign flip in the asymmetry 
at a magnitude larger than the combined uncertainty in 
$|\langle {\cal A}(\phi) \rangle|$, where the statistical error in 
$\langle {\cal A} (\phi) \rangle$ and the systematic error 
$|{\cal A}_{\rm sys}|=0.0009$ have been combined in quadrature. 
Noting Table \ref{tab:ROLR}, and picking the midpoint of the 
$R_i -R_f$ bin with the smallest asymmetry for a $\Delta R$ of 200 pc
we have determined that $R_{\rm flip} = (0.95 \pm 0.03) R_0$.
Our sign flip analysis 
uses the data set of \citet{GHY20} and
implicitly assumes that 
only one effect is operative in the data.  However, we also expect the distorted halo to make some contribution as an overall negative offset to the asymmetry.  This offset, 
determined at larger $R$, is 
small and implies $R_{\rm OLR} > R_{\rm flip}$.  This effect, 
as well as other refinements in our determination of $R_{\rm OLR}$, 
and its implications,  
we plan to analyze in a future paper 
(Hinkel et al. in preparation).

The OLR picture allows for a number of follow-up studies.  First, if the outer Lindblad resonance is responsible 
for the behavior we observe at smaller $R$, we 
should be able to connect the asymmetry to a radial velocity difference
between stars on the left and right of the $\phi = 180^{\circ}$ line.  Studying the axial asymmetry about values of $\phi$ other than $\phi=180^{\circ}$ could also be revealing, as the stellar orbits just within and beyond the OLR have crossing points in $\phi$. Moreover, in this picture we would also not expect to find significant variation of the axial asymmetry with $R$ beyond the Sun's location, which we hope to investigate further, including with the upcoming {\it Gaia} DR3 data set.

\begin{center}
\begin{table}[ht!]
    \centering
    \caption{\label{tab:ROLR}
    Axial asymmetries, N+S, averaged over azimuth angles about the
    anti-center direction up to $|180^{\circ}-\phi|=6^{\circ}$, 
    computed for a wedge of size $\Delta R$ for different choices of starting radius $R_i$, 
    with $R_f=R_i + \Delta R$, to reveal the sign change in the average asymmetry as 
    $R_i -R_f$ changes. We refine the location of the sign flip iteratively by computing the 
    average asymmetry with $R_i$ for smaller $\Delta R$.
    }
    \hskip-2.45cm  
    \begin{tabular}{ |c|c|c|c|c| } 
       \hline
       $R_i - R_f$ (kpc)& $\Delta R$ (kpc)& $\langle {\cal A}(\phi) \rangle$ & $\sigma_{\langle \cal A \rangle}$ & Sign\\ 
       \hline
        7.0 - 7.5& 0.5 & +0.0071 & 0.0012 & +\\ 
       \hline
        7.2 - 7.7& 0.5 & +0.0035 & 0.0011 & +\\ 
       \hline
        7.4 - 7.9& 0.5 & -0.0027 & 0.0011 & -\\ 
       \hline
       \hline
        7.3 - 7.6 & 0.3 & +0.0049 & 0.0012 & +\\ 
       \hline
        7.4 - 7.7& 0.3 & +0.0017 & 0.0012 & +\\ 
       \hline
        7.5 - 7.8& 0.3 & -0.0019 & 0.0012 & -\\ 
       \hline
       \hline
        7.45 - 7.65 & 0.2 & +0.0030 & 0.0013 & +\\ 
       \hline
        7.5 - 7.7& 0.2 & +0.0005 & 0.0013 & +\\ 
       \hline
        7.55 - 7.75& 0.2 & -0.0019 & 0.0012 & -\\ 
       \hline
   \end{tabular}
\end{table}
\end{center}

Finally, we consider whether past effects of the Sgr. dwarf galaxy's collision with the Milky Way could cause an appreciable signal in our data set.  Since the mass used in \citet{GHY20} referred to the present day core of the Sgr dwarf spheroidal, it is interesting to ask if its more massive past self (estimated by \citet{purcell2011sagittarius} to have a mass of roughly $10^{10.5}-10^{11} {\rm M_{\odot}}$) could torque the disk appreciably.  While it is true that Sgr would have applied a significant torque on the galaxy, the axial dispersion velocities of approximately 30 $\rm km\> s^{-1}$ \citep{purcell2011sagittarius} ensure that any asymmetry incurred from the original impact before the significant mass losses of nearly 2 Gyr ago \citep{purcell2011sagittarius} will have been diluted to cover an arc over 50 times larger, reducing the magnitude of any axial asymmetry signal to irrelevance for our volume of space in the present time.  That the vertical motions of stars can be affected by Sagittarius,
suggested in \citet{widrow2012galactoseismology,yanny2013stellar,ferguson2017milky} 
is an entirely different matter.

With regard to the possible influence of the Sgr dwarf tidal stream of the present day we note that a recent impact of Sgr with the disk is very perpendicular to the plane of the disk and at roughly $\phi \approx 180^{\circ}$, thus giving, as we have noted, nearly zero azimuthal torque on the stars in our sample.
This leads us to argue that the LMC, while at a significantly greater distance from the sun than the Sgr tidal stream, induces a much larger left-right asymmetry than Sgr (or the bar if $R \lesssim 8$ kpc). 

\section{Conclusions}

We support
and expand upon the findings of \citet{GHY20} regarding the existence and origin of axial asymmetries in star counts, strongly correlated with mass density in the solar neighborhood, using the Gaia DR2 data set.
After making appropriately conservative cuts, we arrive at a complete, distance-error limited sample of stars, in matched left-right and North-South volumes, which show residual asymmetries in the counts of between 0.5\% to 3\%.  We explore and rule out possible incompleteness due to reddening, lines of sight polluted by regions of high stellar density, significant errors in parallax, magnitude and color limits, and other geometric cuts. 
Based on the estimated azimuthal torques applied to density in the solar neighborhood from Galactic and local group structure, we isolate the Magellanic Cloud system as capable of inducing the largest asymmetry for radii $7.7 < R < 9 $ kpc, as shown in \citet{GHY20}. Recent increases in mass estimates for the LMC \& SMC system \citep{erkal2019total} give this  perturber an outsized influence, compared with earlier estimates.  Going beyond \citet{GHY20}, subdividing the data into three radial bins, we note that the lowest radial bin, $7 < R < 7.7 $ kpc, exhibits a sign flip in 
the asymmetry incompatible with the Magellanic Cloud influence, but consistent with that expected from a bar-induced Outer Lindblad Resonance located slightly beyond $R = 7.6 $ kpc ($R_{\rm flip}=(0.95\pm 0.03) R_{0}$).  While a detailed model which includes several perturbers simultaneously and which considers reflex back-reaction is needed to fully model the detailed asymmetries seen here, we have demonstrated the potential for precision studies of symmetry breaking to constrain and inform our knowledge of the overall mass distribution in and around our Milky Way.

\acknowledgements{
AH thanks the Universities Research Association and the GAANN fellowship for funding.
SG and AH acknowledge partial support from the U.S. Department of Energy under contract DE-FG02-96ER40989.
We also thank the anonymous referee of \citet{GHY20} for helpful insights, which strengthened the investigation of systematics and thank the anonymous referee of this paper for helpful comments that improved the quality of this work.

This document was prepared in part using the resources of Fermi National Accelerator Laboratory (Fermilab), a U.S. Department of Energy, Office of Science, HEP User Facility. Fermilab is managed by the Fermi Research Alliance, LLC (FRA), acting under Contract No. DE-AC02-07CH11359.

This work has made use of data from the European Space Agency (ESA) mission
{\it Gaia} (\url{https://www.cosmos.esa.int/gaia}), processed by the {\it Gaia}
Data Processing and Analysis Consortium (DPAC,
\url{https://www.cosmos.esa.int/web/gaia/dpac/consortium}). Funding for the DPAC
has been provided by national institutions, in particular the institutions
participating in the {\it Gaia} Multilateral Agreement.

}

\bibliography{mybib.bib}

\end{document}